\documentclass[reqno,10pt]{amsart}
\usepackage{amssymb,amsmath,amsthm,amsfonts,color}
\usepackage{subfigure}

\usepackage{mathrsfs}

\usepackage{ulem}
\usepackage{enumerate}
\usepackage{threeparttable}

\usepackage{bm}
\usepackage[left=2.8cm,right=2.8cm,top=2.8cm,bottom=2.8cm]{geometry}
\usepackage{mathtools}
\usepackage{graphicx}
\usepackage[format=hang,labelfont=bf]{caption}

\usepackage[colorlinks=true, pdfstartview=FitV, linkcolor=blue, 
            citecolor=blue, urlcolor=blue]{hyperref}

\allowdisplaybreaks
\numberwithin{equation}{section}
\theoremstyle{plain}
\newtheorem{theorem}{Theorem}[section]

\theoremstyle{definition}

\newtheorem{remark}[theorem]{Remark}

\newtheorem*{theorem*}{Theorem}

\begin{document} 
\title{Discrete-velocity-direction models of BGK-type with minimum entropy: II. Weighted models}

\author{Yihong Chen}
\address{Department of Mathematical Sciences, Tsinghua University\\
    Beijing, 100084, China}
\email{chenyiho20@mails.tsinghua.edu.cn}

\author{Qian Huang*}
\address{Department of Energy and Power Engineering, Tsinghua University\\
    Beijing, 100084, China}
\email{huangqian@tsinghua.edu.cn}

\author{Wen-An Yong*}
\address{Department of Mathematical Sciences, Tsinghua University\\
    Beijing, 100084, China}
\thanks{* Corresponding author}
\email{wayong@tsinghua.edu.cn}


\vskip .2truecm
\begin{abstract}
    In this series of works, we develop a discrete-velocity-direction model (DVDM) with collisions of BGK-type for simulating gas flows, where the molecular motion is confined to some prescribed directions but the speed is still a continuous variable in each orientation.
    In this article, we introduce a weighted function in each orientation when recovering the macroscopic parameters.
    Moreover, the internal molecular degrees of freedom are considered.
    With this weighted DVDM, we develop three submodels by incorporating the discrete velocity method, the Gaussian-extended quadrature method of moments and the Hermite spectral method in each direction.
    These spatial-time submodels are novel multidimensional versions corresponding to the three approaches.
    Numerical tests with a series of 1-D and 2-D flow problems show the efficiency of the weighted DVDM.
\end{abstract}

\maketitle

\normalem

\section{Introduction}
\label{sec:intro}

Gas flows with both continuum and rarefied regimes have found wide applications in areas including lunar-lander-induced dusty plumes \cite{Morris2011}, dynamics of ultrafine particles \cite{Phillips1975} and vacuum technology \cite{Sharipov2007}. The motion of rarefied gases can be properly modelled by kinetic equations describing the time evolution of problem-specific distribution functions \cite{Harris2004,Sharipov2016}. Moreover, the spirit of kinetic theory has been used to describe many interacting systems out of equilibrium, including the airborne aerosols \cite{Friedlander2000,Huang2017} and active matter systems \cite{Bellomo2021}.

Numerical solution of the kinetic equations, including the Boltzmann equation, is rather expensive due to the binary collision terms as well as the high dimensionality of the distribution functions (of space, velocity and other intrinsic properties) \cite{Dimarco2014}.
To overcome such difficulties, various approximations of the kinetic equations have been proposed. Among them, the BGK model is a formal simplification of the Boltzmann equation with the binary collision term  replaced by a relaxation process towards local equilibriums (Maxwellian) \cite{Bhatnagar1954,Welander1954}.
This model retains key properties of the Boltzmann equation, including the conservation laws of mass, momentum and energy, the $H$-theorem, and the correct hydrodynamic limit \cite{Harris2004,Perthame1989,Dimarco2014}.
Moreover, the BGK model becomes so influential that its variants, such as the stochastic particle BGK method \cite{Fei2021} and the BGK-type lattice Boltzmann method \cite{Hou1995,Kruger2017}, have been developed.

Furthermore, there are lasting efforts to seek efficient numerical methods solving the BGK equation \cite{Dimarco2014}. At this point, it should be mentioned that the nonlinearity and non-local dependency of the Maxwellian have posed significant challenges to velocity discretization. The issue is resolved after Mieussens presented an elegant discrete-velocity BGK (DV-BGK) model based on the minimum entropy principle \cite{Mieussens_2000,Mieussens_2000_b}. However, the computational cost raised by high dimensions remains the major obstacle for a direct discretization of the phase space.

On the other hand, the moment methods may be a cure by prescribing the form of the distribution functions. This approach leads to moment closure systems with their hydrodynamic counterpart being the Euler equations. For instance, the well-known Grad’s 13-moment theory \cite{Grad1949} rests on an expansion of the Maxwellian, but it then becomes less reliable far from equilibrium. In contrast, the quadrature-based method of moments (QBMM) \cite{Chalons2017,Fox2008} is free from such limitations at its very heart of model assumptions. Particularly, the Gaussian extended quadrature method of moments (Gaussian-EQMOM) reconstructs the velocity distribution as a summation over several Gaussian functions, and the resultant nonlinear moment system for BGK satisfies the desired structural stability conditions \cite{HLY2020}. However, despite the well-posedness of Gaussian-EQMOM in 1-D velocity space, it is difficult to deduce a multidimensional version of QBMM.

Aiming at a higher-dimensional version of Gaussian-EQMOM, we proposed a discrete-velocity-direction model for the BGK equation by forcing all particles to move in a set of prescribed directions (denoted BGK-DVDM or simply DVDM) in \cite{Huang_2022_arxiv}. The discrete equilibrium is determined with the minimum entropy principle. We have managed to demonstrate the existence, uniqueness and numerical feasibility of thus defined equilibriums. This novel semi-continuous model is flexible enough to incorporate various strategies of removing the 1-D velocity dependence to generate spatial-time models. Remarkably, the extension of Gaussian-EQMOM by the DVDM yields a hyperbolic multidimensional moment closure system \cite{Huang_2022_arxiv}. Moreover, combining the DVDM with the discrete velocity method (DVM) in each direction gives a radial pattern of discrete velocity nodes, rather than the conventional nodes selected on a uniform cubic lattice \cite{Mieussens_2000,Mieussens_2000_b}. Let us mention that the DVDM seems a common practice in solving radiative transfer equations (termed as the discrete ordinates method therein) \cite{Fiveland1987}, while the DVDM for the Boltzmann equation was developed in \cite{ZhangZ2008}.

In this paper, we improve the BGK-DVDM in two aspects.
First, the internal molecular degrees of freedom is included so that more realistic fluid properties can be realized \cite{Chu1965}. 
More importantly, we introduce a weighted function in each orientation when recovering the macroscopic parameters, as opposed to the previous treatment. For the new weighted DVDM, the established properties of the well-behaved discrete equilibrium in \cite{Huang_2022_arxiv} still hold. 
Then the 1-D DVM, Gaussian-EQMOM and a Hermite spectral method \cite{Hu_2020} are applied to generate various DVDM submodels.
Our numerical tests show the significance of the weighted function.

In doing the numerical simulation, we specify proper gas-surface boundary conditions \cite{Sharipov2016} and second-order finite-volume spatial-time schemes \cite{Pareschi_2005} for the new DVDM submodels. The performance of our DVDM submodels is examined with a series of 1-D and 2-D problems covering a wide range of flow regimes, including 2-D Riemann problems and lid-driven cavity flows. The numerical results show that our BGK-DVDM with internal degrees of freedom is a promising multiscale flow solver. It might be worthwhile to study the DVDM with the moment methods further.

The rest of the paper is organized as follows. In Section \ref{sec:models}, we introduce the weighted BGK-DVDM with internal degrees of freedom. Section \ref{sec:submodels} develops three DVDM submodels, including DVD-DVM in Section \ref{sec:dvm}, DVD-EQMOM in Section \ref{sec:dvd_eqmom} and DVD-HSM in Section \ref{sec:dvd_hsm}. 
Boundary conditions for the DVDM submodels are presented in Section \ref{sec:boundary}. Sections \ref{sec:algorithm} and \ref{sec:schemes} contain the algorithms to compute the discrete equilibria and the space-time discretization schemes, respectively.
Numertical results are reported in Section \ref{sec:num_results}.
Finally, some conclusions are given in Section \ref{sec:conclusions}.

\section{Model development}
\label{sec:models}
\subsection{BGK equation with internal degrees of freedom}
\label{sec:BGK}

We start with the BGK equation for the density function  $f=f(t,\bm x, \bm \xi,\bm \zeta)$: 
\begin{equation}\label{eq:1_bgk}
    \partial_t f + \bm \xi \cdot \nabla_{\bm x} f = \frac{1}{\tau} \left( \mathcal{E}[f] - f \right).
\end{equation}
Here, $(\bm x,\bm\xi) \in \mathbb{R}^D \times \mathbb{R}^D$ with $D=2$ or $3$, $\bm \zeta \in \mathbb{R}^L$ represents the possible internal molecular degrees of freedom, and $\tau$ is a characteristic collision time. The internal molecular degrees of freedom determine the specific heat ratio of ideal gases \cite{Guo_2015}. For example, the diatomic gases have two degrees of freedom: rotation and vibration. For planar flows ($D=2$), $\bm \xi$ is just a part of the $3$-dimensional molecule velocity and the rest part is included in $\bm \zeta$.

In the right-hand side (RHS) of Eq.~(\ref{eq:1_bgk}), the local equilibrium state $\mathcal{E}[f] = \mathcal{E}[f](t, \bm x, \bm \xi, \bm \zeta)$ is modeled as
\begin{equation} \label{eq:2_bgk_equil}
    \mathcal{E}[f] = \frac{\rho}{(2\pi \theta)^{(D+L)/2}} \exp \left( -\frac{|\bm \xi - \bm U|^2 + |\bm \zeta|^2}{2\theta} \right).
\end{equation}
Here $|\bm U|$ denotes the Euclidean length of the vector $\bm U$. The classical fluid quantities including density $\rho$, velocity $\bm U$, energy $E$, temperature $\theta$ and pressure $p$ are defined by
\begin{equation} \label{eq:3_macpara}
    \rho = \langle f \rangle, \quad
    \bm U = \frac{\langle \bm \xi f \rangle}{\rho} \in \mathbb{R}^D, \quad
    E = \frac{|\bm U|^2+(D+L)\theta}{2} = \frac{1}{\rho} \left \langle \frac{|\bm \xi|^2 + |\bm \zeta|^2}{2} f \right \rangle, \quad
    p=\rho \theta,
\end{equation}
where the bracket $\langle \cdot \rangle$ is defined as the integral $\langle g(\bm \xi,\bm \zeta) \rangle = \int_{\mathbb{R}^L} \int_{\mathbb{R}^D} g(\bm \xi, \bm \zeta) d \bm \xi d \bm \zeta$ for any reasonable $g(\bm \xi, \bm \zeta)$.
The equilibrium can be rewritten in a concise form
\[
\mathcal{E}[f] = \exp \left( \bm{\alpha}_{eq} \cdot \bm m (\bm \xi, \bm \zeta) \right)
\]
with
\begin{equation} \label{eq:4_m_and_alpha}
    \bm{\alpha}_{eq} = \left( \ln \frac{\rho}{(2\pi \theta)^{(D+L)/2}} - \frac{|\bm U|^2}{2\theta}, \ \frac{\bm U}{\theta}, \ -\frac{1}{\theta} \right)^T \quad\text{and}\quad
    \bm m (\bm \xi, \bm \zeta) = \left( 1, \ \bm \xi, \ \frac{|\bm \xi|^2+|\bm \zeta|^2}{2} \right)^T
\end{equation}
both being $(D+2)$-dimensional real vectors.
This form enlightens us on the model development in later sections.

The equilibrium distribution $\mathcal{E}[f]$ satisfies two important properties. First, $\mathcal{E}[f]$ reproduces the local macroscopic quantities in the same manner as $f$:
\begin{equation} \label{eq:5_equil_mac}
    \left \langle \bm m(\bm \xi, \bm \zeta) \mathcal{E}[f] \right \rangle = \bm \rho := \left( \rho,\ \rho \bm U, \ \rho E \right)^T \in \mathbb{R}^{D+2},
\end{equation}
and thus the BGK equation respects the conservation laws of mass, momentum, and energy. Then, given any $\bm \rho \in \mathbb{R}^{D+2}$ with positive components $\rho$ and $E$, $\mathcal{E}[f]$ is the unique non-negative solution that minimizes the following kinetic entropy 
\begin{equation} \label{eq:6_entropy}
    H[f] = \langle f \ln f - f \rangle
\end{equation}
subject to the constraint $\left \langle \bm m(\bm \xi, \bm \zeta) f \right \rangle = \bm \rho$.

This property has been adapted in both a discrete-velocity model \cite{Mieussens_2000}
and our previous discrete-velocity-direction model for the BGK equation without internal degrees of freedom \cite{Huang_2022_arxiv}.

\subsection{Discrete-velocity-direction models}
\label{sec:dvdm}

A discrete-velocity-direction model (DVDM) based on the BGK equation has been proposed in our previous work \cite{Huang_2022_arxiv}. Our aim here is to enhance the model and extend it to the case with internal molecular degrees of freedom.

The DVDM assumes that the molecule transport is limited to $N$ prescribed directions denoted by $\left\{\bm{l}_m \right\}_{m =1}^N$ with each $\bm{l}_m$ located on the unit sphere $\mathbb{S}^{D-1}$, but the velocity magnitude $\xi \in \mathbb{R}$ in each direction remains continuous.
The directions are selected with the following two requirements.
\begin{enumerate}[(A)]
    \item $(\bm l_1, \dots, \bm l_N) \in \mathbb{R}^{D\times N}$ is of rank $D$ and therefore $N\ge D$.
    \item Each direction $\bm l_m$ and its opposite $-\bm l_m$ belong to $S_m \subset \mathbb{S}^{D-1}$, where the $S_m$'s constitute a disjoint partition of the unit sphere $\mathbb{S}^{D-1} = \bigcup_{m=1}^N S_m$ and each $S_m$ has the same measure.
\end{enumerate}
The equal measure means that the directions are `uniformly distributed'.
For $D=2$, such a partition on $\mathbb{S}^1$ can be realized by setting $\bm l_m = (\cos \gamma_m, \sin \gamma_m)$ and $\gamma_m = \frac{(m-1)\pi}{N}$ or $\frac{(2m-1)\pi}{2N}$, which will be adopted for all numerical tests in this paper.
For $D=3$, the algorithm in \cite{Leopardi_2006} can help to yield such a partition on $\mathbb{S}^2$.

Once the directions are selected, the distribution $f=f(t,\bm x,\bm \xi,\bm \zeta)$ is replaced by $N$ distributions $\left\{f_m(t,\bm x,\xi,\bm \zeta) \right\}_{m=1}^N$ with $\xi \in \mathbb{R}$ and $\bm \zeta \in \mathbb{R}^L$.
The transport velocity for $f_m = f_m(t,\bm x, \xi, \bm \zeta)$ is $\xi \bm l_m$,
and the governing equation for $f_m$ becomes
\begin{equation} \label{eq:7_dvdm_bgk}
    \partial_t f_m + \xi \bm{l}_m \cdot \nabla_{\bm x} f_m = \frac{1}{\tau} (\mathcal{E}_m - f_m), \quad m=1,\dots,N,
\end{equation}
with the local equilibriums $\mathcal{E}_m = \mathcal{E}_m(t,\bm x,\xi,\bm \zeta)$ yet to be determined.

For Eq.~(\ref{eq:7_dvdm_bgk}), we use the weight function $|\xi|^{D-1}$ and define new fluid quantities
\begin{equation} \label{eq:8_macpara_dvdm}
    \begin{aligned}
        \rho &= s \sum_{m=1}^N \int_{\mathbb{R}}\int_{\mathbb{R}^L} f_m |\xi|^{D-1} d\bm\zeta d\xi,\quad
        \rho \bm U= s \sum_{m=1}^N \int_{\mathbb{R}}\int_{\mathbb{R}^L} \xi \bm l_m f_m |\xi|^{D-1} d\bm\zeta d\xi, \\
        \rho E &=  s \sum_{m=1}^N \int_{\mathbb{R}}\int_{\mathbb{R}^L}
        \frac{\xi^2 + |\bm \zeta|^2}{2} f_m |\xi|^{D-1} d\bm\zeta d\xi.
    \end{aligned}
\end{equation}
Here $s$ is half of the measure of $S_m$.

\begin{remark}
    In contrast to our previous model in \cite{Huang_2022_arxiv}, the local equilibrium for the new model Eq.~(\ref{eq:7_dvdm_bgk}) will be evaluated at the just defined fluid quantities computed with the weight function $|\xi|^{D-1}$.
    This weight function is inspired by changing variables from the Cartesian coordinate to polar or spherical coordinates.
    Its introduction is independent of the internal degrees of freedom. 
    Our numerical tests show that this weight function is substantial for correctly reconstructing macroscopic quantities. 
\end{remark}

As for the equilibrium states $\{ \mathcal{E}_m \}_{m=1}^N$ on the RHS of Eq.~(\ref{eq:7_dvdm_bgk}), we require that the following conservation property must be satisfied:
\begin{equation} \label{eq:9_dvdm_equil_consv}
    s \sum_{m=1}^N \int_{\mathbb{R}}\int_{\mathbb{R}^L}  \left(1, \xi \bm{l}_{m}, \frac{\xi^2 + |\bm \zeta|^2}{2} \right) \mathcal{E}_{m} |\xi|^{D-1} d\bm\zeta d\xi = (\rho, \rho \bm U, \rho E).
\end{equation}
This can be viewed as a discrete-velocity-direction analogue of Eq.~(\ref{eq:5_equil_mac}), while $\rho$, $\bm U$, and $E$ are computed with the weighted integrals Eq.~(\ref{eq:8_macpara_dvdm}) based on $f_{m}$.
In this way, we can derive the classical Euler equations by multiplying $1$,
$\xi \bm{l}_{m}$
and $\frac{\xi^2 + |\bm \zeta|^2}{2}$ on both sides of Eq.~(\ref{eq:7_dvdm_bgk}) and taking the weighted integrals; see details in \cite{Huang_2022_arxiv}.

Next we assume that the local equilibrium $\mathcal{E}_m=\mathcal{E}_m(t,\bm x,\xi,\bm \zeta)$ has the variable-separating form
\[
\mathcal{E}_m(t,\bm x,\xi,\bm \zeta) = \mathcal{E}_{tr,m}(t,\bm x,\xi) \mathcal{E}_{in,m}(t,\bm x,\bm \zeta),
\]
which is consistent with the Maxwellian Eq.~(\ref{eq:2_bgk_equil}) of the BGK equation.
The internal part $\mathcal{E}_{in,m}$ is taken to be the same as that in Eq.~(\ref{eq:2_bgk_equil}):
\[
\mathcal{E}_{in,m}(t,\bm x,\bm \zeta) = \frac{1}{\sqrt{(2\pi \theta)^L}} \exp\left( - \frac{|\bm \zeta|^2}{2 \theta} \right).
\]
Notice that the equilibrium temperature $\theta$ is 
\begin{equation}\label{eq:10_rt_eq}
    \theta = \frac{2E - |\bm U|^2}{D+L}
\end{equation}
due to Eq.~(\ref{eq:3_macpara}).
Substituting such an $\mathcal{E}_{in,m}$ into Eq.~(\ref{eq:9_dvdm_equil_consv}), we derive constraints for the transport part $\mathcal{E}_{tr,m}(t,\bm x,\xi)$:
\begin{equation} \label{eq:11_tr_consv}
    s \sum_{m=1}^N \int_{\mathbb{R}} \bm m_{m} \mathcal{E}_{tr,m} |\xi|^{D-1} d\xi
    = \left(\rho, \rho \bm U, \rho \left( E - \frac{L}{2} \theta \right) \right)^T =: \bm \rho_{tr},
\end{equation}
where $\bm m_m(\xi) = \left(1, \xi \bm l_m, \xi^2/2 \right)^T \in \mathbb{R}^{D+2}$.
To determine the transport part, we refer to the minimum entropy property of the Maxwellian in Eq.~(\ref{eq:6_entropy}) and require that $\mathcal{E}_{tr,m}$ minimizes a discrete analogue of the entropy
\begin{equation} \label{eq:12_discEntropy_def}
    H \left[\{g_m \}_{m=1}^N \right] := s \sum_{m=1}^N \int_{\mathbb{R}} \int_{\mathbb{R}^L}  (g_{m} \ln g_{m} - g_{m}) |\xi|^{D-1} d\bm\zeta d\xi
\end{equation}
among all possible 1-D distributions $\left \{g_m(\xi) \ge 0 \right \}_{m=1}^N$ satisfying
\[
s \sum_{m=1}^N \int_{\mathbb{R}} \bm m_{m} g_m |\xi|^{D-1} d\xi
= \bm \rho_{tr}.
\]

For the transport part, we have the following theorem which can be proved with the same argument as that of Theorem 2.1 in our previous work \cite{Huang_2022_arxiv}.

\begin{theorem} \label{thm:equil}
    Given $\bm\rho_{tr}\in \mathbb{R}^{D+2}$ satisfying $0<|\bm \rho_{tr}|<\infty$, if there exists $\{ g_m(\xi) \ge 0\}_{m=1}^N$ such that
    \[
    s \sum_{m=1}^N \int_{\mathbb{R}} \bm m_{m} g_{m} |\xi|^{D-1}  d\xi
    = \bm \rho_{tr},
    \]
    then the discrete kinetic entropy Eq.~(\ref{eq:12_discEntropy_def}) has a unique minimizer $\{\mathcal{E}_{tr,m}\}_{m=1}^N$.
    Moreover, the minimizer has the exponential form
    \[
    \mathcal{E}_{tr,m} = \exp(\bm \alpha \cdot \bm m_m)
    \]
    and $\bm \alpha = (\alpha_0,\hat{\bm{\alpha}}, \alpha_{D+1}) \in \mathbb{R}^{D+1}\times \mathbb{R}^-$ is the unique minimizer of the following convex function
    \begin{equation} \label{eq:13_J_def}
        J(\bm \alpha) := s \sum_{m=1}^N \int_{\mathbb{R}} \exp \left( \bm \alpha \cdot \bm{m}_{m} \right) |\xi|^{D-1} d\xi - \bm \rho_{tr} \cdot \bm \alpha.
    \end{equation}
\end{theorem}

Thanks to this result, the computation of $\mathcal{E}_{tr,m}$ only requires solving $\bm \alpha \in \mathbb{R}^{D+2}$ by minimizing $J(\bm \alpha)$. This is particularly beneficial when a large number of directions are used, i.e., $N \gg D+2$. We will present the algorithm for this optimization problem in later section, which is shown to be highly efficient.

In some cases, it is convenient to rewrite $\mathcal{E}_{tr,m}$ in the form of standard Gaussian distribution
\begin{equation} \label{eq:14_dvdm_gau}
    \mathcal{E}_{tr,m} = \exp(\bm \alpha \cdot \bm{m}_m)
    = \frac{\rho_m}{\sqrt{2\pi \sigma^2}} \exp \left(-\frac{(\xi - u_m)^2}{2\sigma^2} \right),
\end{equation}
and the parameters $\rho_m, \ u_m,$ and $\sigma^2$ are related to $\bm \alpha = (\alpha_0, \hat{\bm \alpha}, \alpha_{D+1})$ as follows:
\begin{equation} \label{eq:15_dvdm_gau_para}
    \sigma^2 = -\frac{1}{\alpha_{D+1}}, \quad
    u_m=\left( \hat{\bm \alpha} \cdot \bm{l}_m \right) \sigma^2, \quad
    \rho_m= \sqrt{2\pi \sigma^2} \exp \left( \alpha_0 + \frac{u_m^2}{2\sigma^2} \right).
\end{equation}

\begin{remark}
    Due to the weighted integral in Eq.~(\ref{eq:8_macpara_dvdm}), $\bm \rho_{tr}$ cannot be expressed, in general, by $\rho_m$, $u_m$ and $\sigma^2$ with the simple algebraic relations
    \[
    \rho \ne s \sum_{m=1}^N \rho_m, \quad
    \rho \bm U \ne s \sum_{m=1}^N \rho_m u_m \bm l_m, \quad
    \rho (|\bm U|^2 + D \theta) \ne s \sum_{m=1}^N \rho_m (u_m^2 + \sigma^2).
    \]
\end{remark}

\section{Spatial-time models}
\label{sec:submodels}

Our BGK-DVDM model  Eq.~(\ref{eq:7_dvdm_bgk}) contains continuous variables $\xi \in \mathbb{R}$ and $\bm \zeta \in \mathbb{R}^L$. 
In this section, we treat these variables to derive spatial-time models with only $t$ and $\bm x$ as continuous variables.
As to $\bm \zeta$, we define 
\[
g_m(t,\bm x,\xi) = \int_{\mathbb{R}^L} f_m(t,\bm x,\xi,\bm \zeta) d \bm \zeta, \quad
h_m(t,\bm x,\xi) = \int_{\mathbb{R}^L} |\bm \zeta|^2 f_m(t,\bm x,\xi,\bm \zeta) d \bm \zeta
\]
and derive from Eq.~(\ref{eq:7_dvdm_bgk}):
\begin{equation} \label{eq:16_dvdm_bgk_gh}
    \left \{
    \begin{aligned}
        & \partial_t g_m + \xi \bm{l}_m \cdot \nabla_{\bm x} g_m = \frac{1}{\tau} \left(\exp(\bm\alpha\cdot\bm m_m) - g_m \right), \\
        & \partial_t h_m + \xi \bm{l}_m \cdot \nabla_{\bm x} h_m = \frac{1}{\tau} \left(L\theta \exp(\bm\alpha\cdot\bm m_m) - h_m \right) 
    \end{aligned}
    \right.
\end{equation}
for $m =1,...,N$.

Next we treat $\xi$ with the following three methods: the discrete-velocity method (DVM) \cite{Mieussens_2000}, the extended quadrature method of moment (EQMOM) \cite{Chalons2017}, and the Hermite spectral method (HSM) \cite{Hu_2020}.

\subsection{Discrete-velocity model}
\label{sec:dvm}

To derive this kind of model, we choose a positive integer $M$, a positive real number $\Delta \xi$, and a real number $\xi_0$, which can vary for different directions.
Set $\xi_{mk}= k\Delta \xi + \xi_0$ for $k=1,...,M$. 
Based on Eq.~(\ref{eq:16_dvdm_bgk_gh}), the discrete-velocity model is
\begin{equation} \label{eq:17_gh_mk}
    \left\{
    \begin{aligned}
        & \partial_t g_{mk} + \xi_{mk} \bm{l}_m \cdot \nabla_{\bm x} g_{mk} = \frac{1}{\tau} (g^{eq}_{mk} - g_{mk}), \\
        & \partial_t h_{mk} + \xi_{mk} \bm{l}_m \cdot \nabla_{\bm x} h_{mk} = \frac{1}{\tau} (h^{eq}_{mk} - h_{mk}),
    \end{aligned}
    \right.
\end{equation}
for $k=1,...,M$ and $m =1,...,N$, where $g_{mk}^{eq}$ and $h_{mk}^{eq}$ need to be determined.

To this end, we first compute
\begin{equation} \label{eq:18_mac_dvdm_dvd}
    \begin{aligned}
        \rho & = s \sum_{m=1}^N \sum_{k=1}^{M} g_{mk} |\xi_{mk}|^{D-1} \Delta \xi, \quad
        \rho \bm U = s \sum_{m=1}^N \sum_{k=1}^{M} \xi_{mk} \bm l_m g_{mk} |\xi_{mk}|^{D-1} \Delta \xi, \quad \\
        \rho E & = s \sum_{m=1}^N \sum_{k=1}^{M} \frac{\xi_{mk}^2 g_{mk} + h_{mk}}{2} |\xi_{mk}|^{D-1} \Delta \xi,
    \end{aligned}
\end{equation}
corresponding to the last model.
With these fluid quantities, $\mathcal{E}_{tr,m}$ in Eq.~(\ref{eq:14_dvdm_gau}) can be derived by finding the minimizer of the convex function $J(\bm \alpha)$ in Eq.~(\ref{eq:13_J_def}) (see detailed algorithms in Section \ref{sec:algorithm}). 

Having $\mathcal{E}_{tr,m}$, we determine the discretized equilibriums $g_{mk}^{eq} $ as the minimizer of the discrete entropy
\[
H_m \left[ \{ u_{mk} \}_{k} \right] = \sum_{k=1}^{M} (u_{mk}\ln u_{mk} - u_{mk} ) |\xi_{mk}|^{D-1} \Delta \xi
\]
among all $\{u_{mk}\ge 0\}_{k=1}^{M}$ satisfying the conservation constraint in the $m$-th direction:
\begin{equation} \label{eq:19_rho_tr_m}
    \sum_{k=1}^{M} \bm m_{mk} u_{mk} |\xi_{mk}|^{D-1} \Delta \xi
    = \int_{\mathbb{R}} \bm m_m |\xi|^{D-1} \mathcal{E}_{tr,m} d\xi \in \mathbb{R}^{D+2}
    =: \bm \rho_{tr,m},
\end{equation}
where we have $\bm m_{mk} = \left( 1, \xi_{mk} \bm l_m, \xi_{mk}^2/2 \right)^T$.
With the argument in \cite{Mieussens_2000}, we can prove that this discretized equilibrium has the form
\begin{equation} \label{eq:20_dis_equi_dvd}
    g_{mk}^{eq} = \exp (\bm \alpha_m \cdot \bm m_{mk}),
\end{equation}
where $\bm \alpha_m \in \mathbb{R}^{D+2}$ is the unique minimizer of the convex function
\[
J_m(\bm \alpha) = \sum_{k=1}^{M} \exp(\bm \alpha \cdot \bm m_{mk}) |\xi_{mk}|^{D-1} \Delta \xi - \bm \alpha \cdot \bm \rho_{tr,m}.
\]
Then we set $h_{mk}^{eq} = L\theta g_{mk}^{eq}$ with $\theta = \frac{2E - |\bm U|^2}{D+L}$ as defined in Eq.~(\ref{eq:10_rt_eq}).

With $g^{eq}_{mk}$ and $h^{eq}_{mk}$ determined above, our discrete-velocity spatial-time model reads as
\[
\left \{
\begin{aligned}
    & \partial_t g_{mk} + \xi_{mk} \bm{l}_m \cdot \nabla_{\bm x} g_{mk} = \frac{1}{\tau} (\exp(\bm \alpha_m \cdot \bm{m}_{mk}) - g_{mk}), \\
    & \partial_t h_{mk} + \xi_{mk} \bm{l}_m \cdot \nabla_{\bm x} h_{mk} = \frac{1}{\tau} (L\theta \exp(\bm \alpha_m \cdot \bm{m}_{mk}) - h_{mk})
\end{aligned}
\right.
\]
for $m =1,...,N$, and $k =1,...,M$.
We close this subsection with two remarks on this model.

\begin{remark}[Computation of $\bm \rho_{tr,m}$]
    Since $\mathcal{E}_{tr,m}$ is of the Gaussian form Eq.~(\ref{eq:14_dvdm_gau}), it is not difficult to verify that
    $\bm \rho_{tr,m} = \rho_{tr,m} (1, u_{tr,m} \bm l_m, E_{tr,m})^T$ in Eq.~(\ref{eq:19_rho_tr_m}) can be computed with the following formulae
    \[
    \begin{aligned}
        \rho_{tr,m} & = \frac{\rho_m u_m}{2}
        P\left(\frac{u_m}{\sqrt{2 \sigma^2} } \right) + \rho_m \sqrt{ \frac{2\sigma^2}{\pi}} \exp\left(- \frac{u_m^2}{2 \sigma^2} \right), \\
        \rho_{tr,m}u_{tr,m} & = \frac{\rho_m (u_m^2 + \sigma^2)}{2} P\left(\frac{u_m}{\sqrt{2 \sigma^2} } \right) + \rho_m u_m \sqrt{ \frac{2 \sigma^2}{\pi}} \exp\left(- \frac{u_m^2}{2 \sigma^2} \right), \\
        2\rho_{tr,m}E_{tr,m} & = \frac{\rho_m u_m(u_m^2 + 3\sigma^2)}{2} P\left(\frac{u_m}{\sqrt{2 \sigma^2} } \right) + \rho_m (u_m^2 + 2\sigma^2) \sqrt{ \frac{2 \sigma^2}{\pi}} \exp\left(- \frac{u_m^2}{2 \sigma^2} \right),
    \end{aligned}
    \]
    where $P(x) := \mathrm{erfc}(-x) - \mathrm{erfc}(x)$ and $\mathrm{erfc}(x) = \frac{2}{\sqrt{\pi}} \int_{x}^{\infty} e^{- \eta^2} d \eta$ is the complementary error function.
    $\rho_m$, $u_m$, and $\sigma^2$ are defined in Eq.~(\ref{eq:15_dvdm_gau_para}).
\end{remark}

\begin{remark}
    Besides the above procedure in deriving a discrete-velocity model, there is another way to close Eq.~(\ref{eq:17_gh_mk}). Indeed, the equilibrium $\left \{ g_{mk}^{eq} \right \}_{m,k}$ in Eq.~(\ref{eq:17_gh_mk}) can be taken as the minimizer of the `total' discrete entropy
    \[
    H \left[\{u_{mk}\}_{m,k} \right] = s \sum_{m=1}^N \sum_{k=1}^{M} ( u_{mk} \ln u_{mk} - u_{mk} ) |\xi_{mk}|^{D-1} \Delta \xi
    \]
    among all $\{u_{mk} \ge 0\}_{m,k}$ satisfying the conservation constraint
    \[
    s \sum_{m=1}^N \sum_{k=1}^{M} \left(1,\xi_{mk} \bm l_m, \frac{|\xi_{mk}^2|}{2}\right)u_{mk} |\xi_{mk}|^{D-1} \Delta \xi = \left(\rho, \rho \bm U, \rho \left(E - \frac{L}{2} \theta \right)\right).
    \]
    In this way, $g_{mk}^{eq}=\exp (\bm \alpha \cdot \bm m_{mk})$ and $\bm \alpha \in \mathbb{R}^{D+1} \times \mathbb{R}^-$ minimizes the convex function
    $$J(\bm \alpha) = s \sum_{m=1}^N \sum_{k=1}^{M} \exp(\bm \alpha \cdot \bm m_{mk}) |\xi_{mk}|^{D-1} \Delta \xi - \bm \alpha \cdot \bm \rho_{tr}.$$
    Here $\bm \rho_{tr}$ is defined as in Eq.~(\ref{eq:11_tr_consv}).
    This treatment is a variant of that for the DVM in \cite{Mieussens_2000}, except that the discrete velocity nodes are chosen radially with a weight function $|\xi_{mk}|^{D-1}$. By contrast, the conventional DVM prefers discrete velocity nodes in a cubic lattice in $\mathbb{R}^D$.
    
    Compared with the DVM approach that solves one (larger-scale) optimization problem, our DVD-DVM requires additional computation of $\mathcal{E}_{tr,m}$, $\bm \rho_{tr,m}$ and $N$ optimization problems. But computing $\mathcal{E}_{tr,m}$ and $\bm \rho_{tr,m}$ are numerically efficient, and minimizing each $J_m(\bm \alpha)$ has a smaller scale than $J(\bm \alpha)$.
    Therefore, the computational cost is acceptable.
\end{remark}

\subsection{Gaussian-EQMOM}
\label{sec:dvd_eqmom}

In this subsection, we apply a method of moment to the BGK-DVDM Eq.~(\ref{eq:16_dvdm_bgk_gh}). The $k$-th velocity moment of $g_m(t,\bm x,\xi)$ and $h_m(t,\bm x,\xi)$ are defined as
\[
M^{[g]}_{m,k}(t,\bm x) = \int_{\mathbb{R}} \xi^k g_m(t,\bm x,\xi) d\xi,
\quad
M^{[h]}_{m,k}(t,\bm x) = \int_{\mathbb{R}} \xi^k h_m(t,\bm x,\xi) d\xi
\]
for $k=0,1,2,...$.
To derive the evolution equations for $M^{[g]}_{m,k}$ and $M^{[h]}_{m,k}$, we integrate the BGK-DVDM Eq.~(\ref{eq:16_dvdm_bgk_gh}) to get
\begin{equation} \label{eq:21_dvdm_mom}
    \begin{aligned}
        & \partial_t M^{[g]}_{m,k} + \bm{l}_m \cdot \nabla_{\bm x} M^{[g]}_{m,k+1} = \frac{1}{\tau} \left(\rho_m \Delta_k(u_m,\sigma^2) - M^{[g]}_{m,k} \right), \\
        & \partial_t M^{[h]}_{m,k} + \bm{l}_m \cdot \nabla_{\bm x} M^{[h]}_{m,k+1} = \frac{1}{\tau} \left( L \rho_m \theta \Delta_k(u_m,\sigma^2) - M^{[h]}_{m,k} \right)
    \end{aligned}
\end{equation}
for $k=0,1,2,...$.
Here $\Delta_k(u,\sigma^2)$ denotes the $k$-th moment of the normalized Gaussian function centered at $u$ with a variance $\sigma^2$.
Eq.~(\ref{eq:21_dvdm_mom}) contains infinitely many equations.

To get a system with finite equations, we resort to the Gaussian-EQMOM method. In this method, it is assumed that the 1-D distribution $g_m(\xi)$ (and $h_m(\xi)$) is a sum of $M$ Gaussian functions \cite{MF2013}:
\begin{equation} \label{eq:22_eqmom_assump}
    \phi_m(\xi) = \sum_{\alpha=1}^{M} \frac{w^{[\phi]}_{m,\alpha}}{\sqrt{2\pi \vartheta^{[\phi]}_m}} \exp \left( -\frac{(\xi-v^{[\phi]}_{m,\alpha})^2}{2 \vartheta^{[\phi]}_m} \right), \quad \text{for }\phi=g \text{ or }h.
\end{equation}
The variance $\vartheta^{[\phi]}_m>0$ is independent on the index $\alpha$. 
With this ansatz, the moments can be expressed as
\begin{equation} \label{eq:23_eqmom_inv}
    M^{[\phi]}_{m,k} = \sum_{\alpha=1}^{M} w^{[\phi]}_{m,\alpha} \Delta_k \left(v^{[\phi]}_{m,\alpha},\vartheta^{[\phi]}_m \right) \quad\text{for } k=0,1,\dots.
\end{equation}
The ansatz above has $2M+1$ parameters $\left( w^{[\phi]}_{m,\alpha}, v^{[\phi]}_{m,\alpha}, \vartheta^{[\phi]}_m \right)$ for $g_m$ or $h_m$. 

To fix these parameters, we reserve the equations in Eq.~(\ref{eq:21_dvdm_mom}) with $k=0,...,2M$ and then solve the first $2M+1$ equations in Eq.~(\ref{eq:23_eqmom_inv}) to express the parameters in terms of the reserved lower moments $M^{[g]}_{m,k}$ and $M^{[h]}_{m,k}$ with $k=0,...,2M$. An algorithm to solve this set of nonlinear algebraic equations can be found in the literature \cite{Chalons2017,MF2013}, which is uniquely solvable in most practical situations \cite{HLY2020}.
In this way, the higher moments $M^{[g]}_{m,2M+1}$ and $M^{[h]}_{m,2M+1}$ in the governing equation of $M^{[g]}_{m,2M}$ and $M^{[h]}_{m,2M}$ can also be expressed in terms of the lower moments
\[
M^{[\phi]}_{m,2M+1} = \sum_{\alpha=1}^{M} w^{[\phi]}_{m,\alpha} \Delta_{2M+1} \left( v^{[\phi]}_{m,\alpha},\vartheta^{[\phi]}_m \right).
\]
Consequently, the equations in Eq.~(\ref{eq:21_dvdm_mom}) with $k=0,...,2M$ are closed.

With the ansatz Eq.~(\ref{eq:22_eqmom_assump}), the macroscopic quantities are naturally computed as
\begin{equation} \label{eq:24_mac_dvd_eqmom}
    \begin{aligned}
        \rho &= s \sum_{m,\alpha} \int_{\mathbb{R}} |\xi|^{D-1} \mathcal{N} \left( \xi; W^{[g]}_{m,\alpha} \right) d\xi, \\
        \rho \bm U &= s \sum_{m,\alpha} \bm l_m \int_{\mathbb{R}} \xi |\xi|^{D-1} \mathcal{N} \left( \xi; W^{[g]}_{m,\alpha} \right) d\xi, \\
        \rho E &= \frac{s}{2} \sum_{m,\alpha} \int_{\mathbb{R}} |\xi|^{D-1} \left[ \xi^2 \mathcal{N} \left( \xi; W^{[g]}_{m,\alpha} \right) + \mathcal{N} \left( \xi; W^{[h]}_{m,\alpha} \right) \right] d\xi,
    \end{aligned}
\end{equation}
where 
$$\mathcal{N} \left(\xi; W^{[\phi]}_{m,\alpha} \right) = \frac{w^{[\phi]}_{m,\alpha}}{\sqrt{2\pi \vartheta^{[\phi]}_m}} \exp \left( -\frac{(\xi-v^{[\phi]}_{m,\alpha})^2}{2 \vartheta^{[\phi]}_m} \right) \quad \text{ and } \quad W^{[\phi]}_{m,\alpha}=\left( w^{[\phi]}_{m,\alpha}, v^{[\phi]}_{m,\alpha}, \vartheta^{[\phi]}_m \right) \in \mathbb{R}^{3}.$$
Notice that due to the weight function $|\xi|^{D-1}$ in Eq.~(\ref{eq:24_mac_dvd_eqmom}), we generally have
\[
\rho \ne s \sum_m M^{[g]}_{m,0}, \quad
\rho \bm U \ne s \sum_m \bm l_m M^{[g]}_{m,1}, \quad
\rho E \ne s \sum_m \frac{1}{2} \left( M^{[g]}_{m,2}+M^{[h]}_{m,0} \right).
\]

Eqs.~(\ref{eq:21_dvdm_mom},\ref{eq:23_eqmom_inv}-\ref{eq:24_mac_dvd_eqmom}) make up a spatial-time model by incorporating Gaussian-EQMOM into the BGK-DVDM Eq.~(\ref{eq:16_dvdm_bgk_gh}).
This model, denoted as DVD-EQMOM, is a convenient multidimensional version of quadrature-based method of moments, which seems better understood than those in \cite{Chalons2017,MF2013}.
Moreover, the moment system is hyperbolic, indicating a well-posed extension of the EQMOM. The proof is similar to that in our previous work \cite{Huang_2022_arxiv} for the BGK equation without internal degrees of freedom. It mainly relies on Ref. \cite{HLY2020}, where the hyperbolicity of the 1-D EQMOM was thoroughly analyzed.

\subsection{Hermite spectral method}
\label{sec:dvd_hsm}

In this subsection we treat the continuous variable $\xi$ with the Hermite spectral method (HSM) proposed in \cite{Hu_2020}. In this method, it is assumed that the distribution $\phi = \phi(t,\bm x, \xi)$ is a truncation
\begin{equation} \label{eq:25_hsm_series}
    \phi(t,\bm x,\xi) = \sum\limits_{k=0}^{M-1} \phi_k(t,\bm x) \mathcal{H}_k^{[\bar u,\bar\theta]}(\xi)
\end{equation}
of a series with the basis function 
\[
\mathcal{H}_n^{[\bar u,\bar\theta]}(\xi) = \bar\theta^{-n/2} H_n\left(\frac{\xi-\bar u}{\sqrt{\bar\theta}}\right) \frac{1}{\sqrt{2\pi\bar\theta}}e^{- \frac{(\xi-\bar u)^2}{2 \bar\theta}}.
\]
Here $M$ is a given integer, 
\[
H_n(x) = (-1)^n e^{\frac{x^2}{2}} \left( \frac{d^n}{dx^n} e^{- \frac{x^2}{2}}\right)
\]
is the $n$th-order Hermite polynomial, and $\bar u$, $\bar \theta$ are two constant parameters.
In this paper, we always set $\bar u =0$ and determine $\bar \theta$ by the initial flow condition.

Due to the orthogonality of the Hermite polynomials:
\[
\int_{\mathbb{R}} \mathcal{H}_n^{[\bar u,\bar\theta]}(\xi)
\frac{\bar\theta^{m/2}}{m!} H_m\left(\frac{\xi-\bar u}{\sqrt{\bar\theta}}\right) d\xi = \delta_{nm},
\]
the coefficient $\phi_k(t,\bm x)$ in Eq.~(\ref{eq:25_hsm_series}) can be uniquely determined as 
\[
\phi_k(t,\bm x) = \int_{\mathbb{R}} \phi(t,\bm x,\xi) \frac{\bar\theta^{k/2}}{k!} H_k\left(\frac{\xi-\bar u}{\sqrt{\bar\theta}}\right) d\xi.
\]
Thus the $M$-truncation of $\phi$ is fully determined.

To incorporate the HSM with the BGK-DVDM Eq.~(\ref{eq:16_dvdm_bgk_gh}), we set 
\[
\begin{aligned}
    \phi^{[g]}_{m,k}(t,\bm x) = \int_{\mathbb{R}} g_m(t,\bm x,\xi) |\xi|^{D-1} \frac{\bar\theta^{k/2}}{k!} H_k\left(\frac{\xi-\bar u}{\sqrt{\bar\theta}}\right) d\xi, \\
    \phi^{[h]}_{m,k}(t,\bm x) = \int_{\mathbb{R}} h_m(t,\bm x,\xi) |\xi|^{D-1} \frac{\bar\theta^{k/2}}{k!} H_k\left(\frac{\xi-\bar u}{\sqrt{\bar\theta}}\right) d\xi.
\end{aligned}
\]
Then we multiply the both sides of Eq.~(\ref{eq:16_dvdm_bgk_gh}) with $|\xi|^{D-1} \frac{\bar\theta^{k/2}}{k!} H_k \left(\frac{\xi-\bar u}{\sqrt{\bar\theta}} \right)$ for $k=0,...,M-1$ and integrate over $\xi \in \mathbb{R}$ to obtain
\begin{equation} \label{eq:26_dvd_hsm}
    \partial_t \bm\Phi_m + \mathcal{A} \bm l_m \cdot \nabla_{\bm x} \bm\Phi_m = \frac{1}{\tau} ( \bm\Phi_m^{eq} - \bm\Phi_m ).
\end{equation}
Here $ \bm\Phi_m = (\phi_{m,0},...,\phi_{m,M-1})^T \in \mathbb{R}^M$ with $\phi_{m,k} = \phi_{m,k}^{[g]}$ or $\phi_{m,k}^{[h]}$, the constant matrix $\mathcal{A} \in \mathbb{R}^{M\times M}$ is tridiagonal \cite{Hu_2020}:
\begin{equation}\label{eq:27_HSM_A}
    \mathcal{A} = \left(
    \begin{matrix}
        \bar u      & 1         &        &            &     \\
        \bar\theta  & \bar u    & 2      &            &     \\
        & \bar\theta& \ddots & \ddots     &     \\
        &           & \ddots & \ddots     & M-1 \\
        &           &        & \bar\theta & \bar u
    \end{matrix}
    \right),
\end{equation}
and the equilibrium $\bm\Phi_m^{eq} = ( \phi^{eq}_{m,0},...,\phi^{eq}_{m,M-1} )^T$ has components
\begin{equation} \label{eq:28_phi_eq_mk}
    \phi^{eq}_{m,k} = \int_{\mathbb{R}} \frac{\bar\theta^{k/2}}{k!} H_k \left(\frac{\xi-\bar u}{\sqrt{\bar\theta}} \right) \phi_m^{eq} d \xi \quad \text{for }\ \phi_m^{eq} = \mathcal{E}_{tr,m}|\xi|^{D-1} \ \text{ or }\ L \theta \mathcal{E}_{tr,m}|\xi|^{D-1}.
\end{equation}
The corresponding macroscopic quantities are computed as
\[
\begin{aligned}
    \rho &= s \sum_{m=1}^N \phi^{[g]}_{m,0}, \quad
    \rho \bm U = s \sum_{m=1}^N \bm l_m \left( \phi^{[g]}_{m,1} + \bar{u} \phi^{[g]}_{m,0} \right), \\
    \rho E &= s \sum_{m=1}^N \frac{1}{2} \left( 2\phi^{[g]}_{m,2} + 2\bar{u}\phi^{[g]}_{m,1} + (\bar{\theta} + \bar{u}^2)\phi^{[g]}_{m,0} + \phi^{[h]}_{m,0} \right).
\end{aligned}
\]
The equations in Eq.~(\ref{eq:26_dvd_hsm}) constitute our third kind of models, denoted as DVD-HSM.

We end this subsection with details on computing $\phi_{m,k}^{eq}$ in Eq.~(\ref{eq:28_phi_eq_mk}) when $\bar{u}=0$. Clearly, we only need to consider $\phi^{eq}_m = \mathcal{E}_{tr,m}|\xi|^{D-1}$ with $\mathcal{E}_{tr,m}$ given in Eq.~(\ref{eq:14_dvdm_gau}). 
Moreover, only the 2-D case is presented because for $D=3$ the weight function has a simpler expression $|\xi|^{D-1}=\xi^2$ and therefore the 3-D case is easier to handle.
To simplify the notation, we set
\[
\begin{aligned}
    a_k &:= \phi_{m,k}^{eq} = \int_{\mathbb{R}} \frac{\bar\theta^{k/2}}{k!} H_k \left(\frac{\xi}{\sqrt{\bar\theta}} \right) \mathcal{E}_{tr,m}(\xi) |\xi| d \xi, \\
    b_k &:= \int_{\mathbb{R}} \frac{\bar\theta^{k/2}}{k!} H_k \left(\frac{\xi}{\sqrt{\bar\theta}} \right) \mathcal{E}_{tr,m}(\xi) \mathrm{sgn}(\xi) d \xi,
\end{aligned}
\]
for $k = 0,1,...$. It is not difficult to see from the recursive formula of Hermite polynomials \cite{Hu_2020}
\[
H_0(x)=1,\quad H_1(x)=x,\quad
H_{n+1}(x) = x H_n(x) - n H_{n-1}(x)
\]
and the relation $|\xi|=\xi \mathrm{sgn}(\xi)$ that
\[
a_0 = b_1 \quad \text{and} \quad
a_k = (k+1) b_{k+1} + \bar \theta b_{k-1} \quad \text{for} \quad k \ge 1.
\]
Thus, it suffices to compute $b_k$. Write
\[
b_k = \int_0^{+\infty} \frac{\bar\theta^{k/2}}{k!} H_k \left(\frac{\xi}{\sqrt{\bar\theta}} \right) \mathcal{E}_{tr,m}(\xi) d \xi
- \int_{-\infty}^0 \frac{\bar\theta^{k/2}}{k!} H_k \left(\frac{\xi}{\sqrt{\bar\theta}} \right) \mathcal{E}_{tr,m}(\xi) d \xi =: b_k^+ - b_k^-.
\]
With the recursive formula, we can obtain
\[
\begin{aligned}
    b_{k+1}^+ = & \int_0^{+\infty} \frac{\bar\theta^{(k+1)/2}}{(k+1)!} \left[ \frac{\xi}{\sqrt{\bar\theta}} H_k \left(\frac{\xi}{\sqrt{\bar\theta}} \right) - k H_{k-1} \left(\frac{\xi}{\sqrt{\bar\theta}} \right) \right] \mathcal{E}_{tr,m}(\xi) d \xi \\
    = & \int_0^{+\infty} \frac{\bar\theta^{k/2}}{(k+1)!} \xi H_k \left(\frac{\xi}{\sqrt{\bar\theta}} \right)  \mathcal{E}_{tr,m}(\xi) d \xi - \frac{\bar \theta}{k+1} b_{k-1}^+ \\
    = & \int_0^{+\infty} \frac{\bar\theta^{k/2}}{(k+1)!} (\xi - u_m) H_k \left(\frac{\xi}{\sqrt{\bar\theta}} \right)  \mathcal{E}_{tr,m}(\xi) d \xi + \frac{u_m}{k+1} b_{k}^+ - \frac{\bar \theta}{k+1} b_{k-1}^+.
\end{aligned}
\]
Note that $\frac{d}{d \xi} \mathcal{E}_{tr,m}(\xi) = - \frac{\xi - u_m}{\sigma^2} \mathcal{E}_{tr,m} (\xi)$ and $H_n'(x) = n H_{n-1}3(x)$. Using the integration by parts gives
\[
\begin{aligned}
    b_{k+1}^+ = & -\int_0^{+\infty} \frac{\sigma^2 \bar\theta^{k/2}}{(k+1)!} H_k \left(\frac{\xi}{\sqrt{\bar\theta}} \right)  d\mathcal{E}_{tr,m}(\xi) + \frac{u_m}{k+1} b_{k}^+ - \frac{\bar \theta}{k+1} b_{k-1}^+ \\
    = & \frac{\sigma^2 \bar\theta^{k/2}}{(k+1)!} H_k (0) \mathcal{E}_{tr,m}(0)
    + \int_0^{+\infty} \frac{\sigma^2 \bar\theta^{(k-1)/2}}{(k+1)!} k H_{k-1} \left(\frac{\xi}{\sqrt{\bar\theta}} \right) \mathcal{E}_{tr,m}(\xi) d \xi
    + \frac{u_m b_{k}^+ - \bar \theta b_{k-1}^+}{k+1} \\
    = & \frac{\sigma^2 \bar\theta^{k/2}}{(k+1)!} H_k (0) \mathcal{E}_{tr,m}(0) + \frac{u_m b_{k}^+ + (\sigma^2 - \bar \theta) b_{k-1}^+}{k+1}.
\end{aligned}
\]
A similar computation can be done for $b_k^-$ and finally we get the following recursive formula
\[
b_{k+1} = \frac{2\sigma^2 \bar\theta^{k/2}}{(k+1)!} H_k (0) \mathcal{E}_{tr,m}(0) + \frac{u_m b_{k} + (\sigma^2 - \bar \theta) b_{k-1}}{k+1}.
\]
Additionally, a direct computation gives 
\[
b_0 = \frac{1}{2}  \left( \mathrm{erfc}\left( \frac{u_m }{ \sqrt{2 \sigma^2}} \right) - \mathrm{erfc}\left( \frac{u_m }{ \sqrt{2 \sigma^2}} \right) \right), \quad b_1 = u_m b_0 + 2 \sigma^2 \mathcal{E}_{tr,m} (0).
\]

\subsection{Brief summary of the models}

\begin{figure}[htbp]
    \centering
    \includegraphics[height=5.2cm]{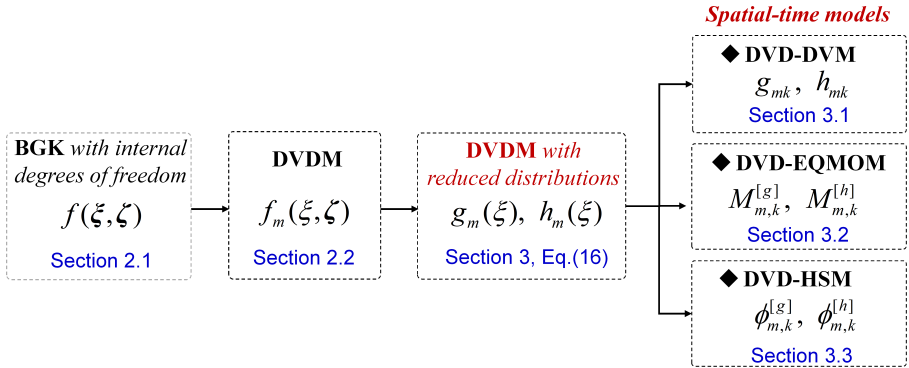}
    \caption{Model hierarchy of the DVDM. The variable and the equilibrium state are shown in each model (block).} \label{fig:models}
\end{figure}

Fig.~\ref{fig:models} presents a brief summary and the hierarchy of the several models proposed up to now.
The original BGK equation with internal molecular degrees of freedom is reviewed in Section~\ref{sec:BGK}. The DVDM assumes that the particles move in $N$ fixed orientations, leading to the model Eq.~(\ref{eq:7_dvdm_bgk}) for $f_m$.
Then, Section~\ref{sec:submodels} develops three spatial-time models by eliminating the continuous variables $\xi$ and $\bm \zeta$, including DVD-DVM in Section~\ref{sec:dvm}, DVD-EQMOM in Section~\ref{sec:dvd_eqmom}, and DVD-HSM in Section~\ref{sec:dvd_hsm}.
For these models, the boundary conditions and numerical schemes need to be specified before practical flow simulations.

\section{Boundary conditions}
\label{sec:boundary}

Let $\Omega \subset \mathbb{R}^{D}$ be the computational domain and denote by $\bm n = \bm n(\bm x)$ the outward unit normal vector of the boundary $\partial \Omega$ at $\bm x$.
Two types of boundary conditions are considered in this paper.
The first one is the Neumann condition $\bm n \cdot \nabla_{\bm x} \phi = 0$ with $\phi$ representing any unknown variables in the DVDM submodels (see Fig.~\ref{fig:models}). 

The second one is the solid wall conditions. 
For simplicity, let the boundary velocity $\bm U_w $ at $\bm x_w \in \partial \Omega$ be perpendicular to $\bm n=\bm n(\bm x_w)$.
For the original BGK equation, the boundary distribution $f(t,\bm x_w, \bm\xi, \bm \zeta)$ for reflecting particles, i.e. $\bm \xi \cdot \bm n < 0$, should be given by the distribution of outgoing particles, i.e. $\bm \xi \cdot \bm n > 0$. Two specific boundary conditions are the diffuse-scattering law and the bounce-back rule (also termed specular-reflection law) \cite{Naris2005,Guo_2013}.
The first one assumes that the distribution of reflecting particles is a Maxwellian:
\begin{equation} \label{eq:diffuse_be}
    f(t,\bm x_w,\bm \xi, \bm \zeta) = \sqrt{\frac{2\pi}{\theta_w}}\frac{j(t,\bm x_w)}{\sqrt{(2 \pi \theta_w)^{D+L}}} \exp \left( - \frac{|\bm \xi - \bm U_w|^2 + |\bm \zeta|^2}{2 \theta_w} \right), \quad \bm \xi \cdot \bm n < 0,
\end{equation}
where $\theta_w$ is the boundary temperature at $\bm x_w$ and $j(t, \bm x_w)$ is the outward-flowing mass flux defined by
\[
j(t,\bm x_w) = \int_{\mathbb{R}^L} \int_{\bm \xi \cdot \bm n > 0} \bm n \cdot \bm \xi f(t,\bm x_w,\bm \xi, \bm \zeta) d \bm \xi d \bm \zeta.
\]
This condition ensures no particle penetration through the boundary.
The bounce-back rule is widely used in the lattice Boltzmann method \cite{Ladd_1994}. It reads as
\[
f(t,\bm x_w,\bm \xi,\bm \zeta) = f(t,\bm x_w, - \bm \xi,\bm \zeta) + 2 \rho_w(t,\bm x_w) \mathcal{E}(\bm \xi, \bm \zeta) \frac{\bm \xi \cdot \bm U_w}{\theta_w}, \quad \bm \xi \cdot \bm n < 0.
\]
Here 
\[
\mathcal{E}(\bm \xi, \bm \zeta) = \frac{1}{\sqrt{(2\pi \theta_w)^{D+L}}} \exp \left( - \frac{|\bm \xi|^2 + |\bm \zeta|^2}{2 \theta_w} \right)
\]
and
\[
\rho_w (t,\bm x_w) = \frac{ 2 \int_{\mathbb{R}^L} \int_{\bm \xi \cdot \bm n > 0} f(t,\bm x_w,\bm \xi,\bm \zeta) d \bm \xi d \bm \zeta }
{ 1 - \frac{2}{\theta_w} \int_{\mathbb{R}^L} \int_{\bm \xi \cdot \bm n < 0} \bm \xi \cdot \bm U_w\mathcal{E}(\bm \xi,\bm \zeta) d \bm \xi d \bm \zeta}.
\]
This condition ensures that the macroscopic velocity $\bm U(t,\bm x_w)$ equals $\bm U_w$. Since $\bm U_w$ is assumed to be perpendicular to $\bm n$, $\rho_w$ is simplified as
\[
\rho_w(t,\bm x_w) = 2 \int_{\mathbb{R}^L} \int_{\bm \xi \cdot \bm n>0} f(t,\bm x_w, \bm \xi, \bm \zeta) d\bm \xi d \bm \zeta.
\]

We now illustrate how these kinetic boundary conditions are adapted to the new DVDM submodels in Section~\ref{sec:submodels}. The main idea is to replace the integrals above by proper discrete sums.

For the DVD-DVM in Subsection~\ref{sec:dvm}, the diffuse-scattering law is converted to 
\[
\begin{aligned}
    g_{mk}(t,\bm x_w) & = \sqrt{\frac{2\pi}{\theta_w}} j(t,\bm x_w) \mathcal{E}_{tr,mk}[\bm U_w, \theta_w], \quad &\xi_k \bm l_m \cdot \bm n < 0, \\
    h_{mk}(t,\bm x_w) & = L \theta_w g_{mk}(t,\bm x_w), \quad &\xi_k \bm l_m \cdot \bm n < 0,
\end{aligned}
\]
with
\[
j(t,\bm x_w) = s \sum_{m=1}^N \sum_{k=1}^M \xi_k \bm l_m \cdot \bm n g_{mk}(t,\bm x_w) |\xi_k|^{D-1} \Delta \xi \bm 1_{\xi_k \bm l_m \cdot \bm n > 0},
\]
and 
$\mathcal{E}_{tr,mk}[\bm U_w, \theta_w]$ the discrete equilibrium defined in Eq.~(\ref{eq:20_dis_equi_dvd}) with density $1$, velocity $\bm U_w$, and temperature $\theta_w$. On the other hand, we assume that $\{\xi_k\}_{k=1}^M$ satisfies $\xi_k = -\xi_{M+1-k}$ for $k=1,...,M$ to apply the bounce-back rule to the DVD-DVM. With this assumption, the discrete-velocity version of bounce-back rule becomes
\[
\begin{aligned}
    g_{mk}(t,\bm x_w) & = g_{m,M+1-k}(t,\bm x_w) + 2 \rho_w(t,\bm x_w) \mathcal{E}_{tr,mk}[\bm 0, \theta_w] \frac{\xi_k \bm l_m \cdot \bm U_w}{\theta_w}, \quad &\xi_k \bm l_m \cdot \bm n < 0, \\
    h_{mk}(t,\bm x_w) & = h_{m,M+1-k}(t,\bm x_w) + 2 \rho_w(t,\bm x_w) L \theta_w \mathcal{E}_{tr,mk}[\bm 0, \theta_w] \frac{\xi_k \bm l_m \cdot \bm U_w}{\theta_w}, \quad &\xi_k \bm l_m \cdot \bm n < 0,
\end{aligned}
\]
where
\[
\rho_w (t,\bm x_w) = \frac{ 2 s \sum_{m=1}^N \sum_{k=1}^M g_{mk}(t,\bm x_w) |\xi_k|^{D-1} \Delta \xi \bm 1_{\xi_k \bm l_m \cdot \bm n > 0} }
{ 1 - \frac{2}{\theta_w} s \sum_{m=1}^N \sum_{k=1}^M \xi_k \bm l \cdot \bm U_w \mathcal{E}_{tr,mk}[\bm 0, \theta_w] |\xi_k|^{D-1} \Delta \xi \bm 1_{\xi_k \bm l_m \cdot \bm n < 0} }.
\]

For the DVD-EQMOM in Subsection~\ref{sec:dvd_eqmom}, only the diffuse-scattering law is used, which reconstructs the velocity distributions of reflecting particles as
\[
\begin{aligned}
    g_{m}(t,\bm x_w, \xi) & = \sqrt{\frac{2\pi}{\theta_w}} j(t,\bm x_w) \mathcal{E}_{tr,m}[(1,\bm U_w, \theta_w)], \quad &\xi \bm l_m \cdot \bm n < 0, \\
    h_{m}(t,\bm x_w, \xi) & = L \theta_w g_{m} (t,\bm x_w, \xi), \quad &\xi \bm l_m \cdot \bm n < 0,
\end{aligned}
\]
with
\[
j(t,\bm x_w) = s \sum_{m=1}^N \int_{\xi \bm l_m \cdot \bm n > 0} \xi \bm l_m \cdot \bm n |\xi|^{D-1} \sum_{\alpha=1}^{M} \mathcal{N} \left( \xi; W^{[g]}_{m,\alpha} \right) d \xi,
\]
and $\mathcal{E}_{tr,m}[\bm U_w, \theta_w]$ the discrete equilibrium defined in Eq.~(\ref{eq:14_dvdm_gau}) with density $1$, velocity $\bm U_w$, and temperature $\theta_w$. Other notations follow the definitions in Section~\ref{sec:dvd_eqmom}.
The moments on the boundary can then be evaluated as 
$$
M_{m,k}^{[\phi]}(t,\bm x_w) = \int_{\{\xi \bm l_m \cdot \bm n < 0\} \bigcup \{\xi \bm l_m \cdot \bm n > 0\}} \xi^k \phi_m(t,\bm x_w, \xi) d\xi
$$
for $\phi_m=g_m$ or $h_m$. Note that the integrand takes different forms in the two sets.

For the DVD-HSM in Subsection \ref{sec:dvd_hsm}, further boundary conditions are left for future work.

\section{Algorithms}
\label{sec:algorithm_and_scheme}

\subsection{Algorithm for the discrete equilibrium}
\label{sec:algorithm}

Solving the discrete equilibrium $\mathcal{E}_{tr,m}$ defined in Eq.~(\ref{eq:11_tr_consv}) out of a known $\bm \rho_{tr}$ is necessary for all DVDM submodels in Section~\ref{sec:submodels}. Theorem~\ref{thm:equil} indicates that all we need is an $\bm \alpha \in \mathbb{R}^{D+1} \times \mathbb{R}^-$ that minimizes the convex function $J(\bm \alpha)$ in Eq.~(\ref{eq:13_J_def}).
The gradient descent method was used in our previous work \cite{Huang_2022_arxiv}, while we use the BFGS quasi-Newton method \cite{Broyden_1970} in this work.

\begin{figure}[htbp]
    \centering
    \includegraphics[height=5cm]{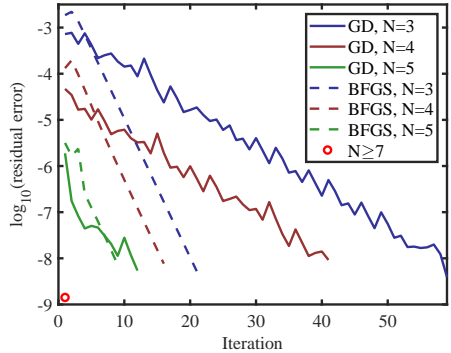}
    \caption{Performance of the BFGS quasi-Newton method and the gradient descend (GD) method for $D=2$, $\rho=1$, $\bm U=(0.4,0.8)^T$, and $\theta=0.8$. }\label{fig:1}
\end{figure}

Fig.~\ref{fig:1} presents the performance of the BFGS quasi-Newton method and the gradient descend (GD) method for $D=2$. Here we set $\rho=1$, $\bm U=(0.4,0.8)^T$, and $\theta=0.8$.
The discrete directions are chosen as $\left\{ \bm l_m = \left( \cos\frac{(m-1)\pi}{N}, \sin\frac{(m-1)\pi}{N} \right)^T \right\}_{m=1}^N$.
The BFGS method requires much less iteration steps to converge for $N \le 5$. Notably, when $N\ge 7$, the initial value $\bm \alpha_{eq}$ is so close to the minimizer $\bm \alpha$ that only one step of iteration leads to convergence. Therefore, the computation of discrete equilibrium in the DVDM is numerically efficient.

\subsection{Numerical schemes}
\label{sec:schemes}

In this subsection we present some numerical schemes to solve the DVDM submodels proposed before. 
Recall that both the DVD-DVM and DVD-HSM can be written in a unified form as
\begin{equation}  \label{eq:unified_goveq}
    \partial_t \bm\Phi_m + \mathcal{A} \bm l_m \cdot \nabla_{\bm x} \bm\Phi_m = \frac{1}{\tau} (\bm\Phi_m^{eq} - \bm\Phi_m) =: \bm\Omega_m,\quad m=1,...,N.
\end{equation}
For the DVD-DVM, we have $\bm \Phi_m = (\phi_{m1},...,\phi_{mM})^T \in \mathbb{R}^M$ with $\phi_{mk} = g_{mk}$ or $h_{mk}$, and the matrix $\mathcal{A} = {\rm diag} \{ \xi_1,...,\xi_M\}$.
For the DVD-HSM, $\bm \Phi_m= (\phi_{m,0},...,\phi_{m,M-1})^T \in \mathbb{R}^M$ and $\mathcal{A}$ are defined in Eqs.~(\ref{eq:26_dvd_hsm} \& \ref{eq:27_HSM_A}).

For a time discretization of Eq.~(\ref{eq:unified_goveq}), the implicit-explicit Runge-Kutta (IMEX-RK) schemes \cite{Pareschi_2005} can be applied. Here we only use a second-order scheme denoted by SSP2. It is characterised by a double tableau \cite{Pareschi_2005}
\[
\begin{array}{c|cc}
    0 & 0 & 0 \\
    1 & 1 & 0 \\
    \hline
    & 1/2 & 1/2
\end{array}
\quad
\begin{array}{c|cc}
    \gamma & \gamma & 0 \\
    1 - \gamma & 1-2 \gamma & \gamma\\
    \hline
    & 1/2 & 1/2
\end{array},
\quad
\gamma = 1 - \frac{1}{\sqrt{2}}.
\]
Although the source term $\bm \Omega_m$ is implicitly discretized, its relaxation structure renders a well-known way to solve the equations explicitly (see e.g. \cite{Guo_2013,Guo_2015}).
The convection term is treated with 
the third-order energy stable WENO (ES-WENO) scheme \cite{Yamaleev_2009}.
For the DVD-DVM, the Godunov flux \cite{Pareschi_2005} is adopted, 
while the HLL flux \cite{Harten_1982,Hu_2020} is used for the DVD-HSM.

On the other hand, for the DVD-DVM, Eq.~(\ref{eq:unified_goveq}) can also be discretized with upwind schemes of first-order accuracy, which renders a easier way to treat the boundary conditions. An implicit discretization for the collision term can be treated similarly as in the IMEX-RK scheme. 

Finally for the DVD-EQMOM, the $M^{[g]}_{m,k}$-equation in Eq.~(\ref{eq:21_dvdm_mom}) can be approximated by the 2-D upwind scheme
\begin{equation} \label{eq:sch_dvdeqmom}
    \begin{aligned}
        M^{[g],n+1}_{m,k,ij} = & M^{[g],n}_{m,k,ij} - \frac{\Delta t}{\Delta x} \bm l_m \cdot \bm e_1 \left( \mathcal{G}^n_{m,k+1,i+\frac{1}{2},j} - \mathcal{G}^n_{m,k+1,i-\frac{1}{2},j} \right) \\
        -& \frac{\Delta t}{\Delta y} \bm l_m \cdot \bm e_2 \left( \mathcal{G}^n_{m,k+1,i,j+\frac{1}{2}} - \mathcal{G}^n_{m,k+1,i,j-\frac{1}{2}} \right)
        + \frac{\Delta t}{\tau} \left( M^{[g],n}_{\mathcal{E} m,k,ij} - M^{[g],n+1}_{m,k,ij} \right)
    \end{aligned}
\end{equation}
with a partially implicit collision term. The $M^{[h]}_{m,k}$-equation is treated similarly. 
Here the fluxes 
\begin{equation} \label{eq:flux_mom_2}
    \mathcal{G}^n_{m,k+1,i+\frac{1}{2}} =\left \{
    \begin{aligned}
        \int_0^{\infty} \xi^{k+1} g_{m,ij}^n d\xi + \int_{-\infty}^0 \xi^{k+1} g_{m,i+1,j}^n d\xi, \quad &\text{if } \bm l_m \cdot \bm e_1 >0, \\
        \int_0^{\infty} \xi^{k+1} g_{m,i+1,j}^n d\xi + \int_{-\infty}^0 \xi^{k+1} g_{m,ij}^n d\xi, \quad &\text{if } \bm l_m \cdot \bm e_1 <0
    \end{aligned}
    \right.
\end{equation}
are the same as those in \cite{Chalons2017,MF2013}.
The moments 
\[
M^{[g],n}_{\mathcal{E}_{m,k,ij}} = \rho^n_{m,ij} \Delta_k \left( u^n_{m,ij}, (\sigma^2)^n_{ij} \right)
\]
correspond to the equilibrium state, where $\Delta_k(u,\sigma^2)$ is defined in Section \ref{sec:dvd_eqmom}. The equilibrium state parameters $\rho^n_{m,ij}, \ u^n_{m,ij}$ and $(\sigma^2)^n_{ij}$ are obtained by solving the local equilibrium Eq.~(\ref{eq:14_dvdm_gau}).


\section{Numerical results}
\label{sec:num_results}

In this section, we present the results of some numerical tests based on the discretizations of the previous DVDM submodels. The tests only involve planar flows ($D=2$).

\subsection{1-D Riemann Problems}
\label{sec:1D_Riemann_prob}

We start with 1-D Riemann problems. Assume no internal degrees of freedom ($L=0$). The Riemann initial data of the fluid quantities read as \cite{Fox2008}:
\[
\rho(0,x) = \left\{
\begin{aligned}
    & 3.093, & x<0,\\
    & 1, & x>0,
\end{aligned}
\right.
\quad
\bm U(0,x) = \bm 0,
\quad
\theta(0,x) = 1.
\]
Both the continuum (infinitely fast collision limit $\tau=0$) and free-molecular (no collision limit $\tau=\infty$) regimes are considered. The theoretical solutions for both cases can be found in \cite{Lora_2013} and \cite{Guo_2015}.
The 1-D physical domain $[-0.5,0.5]$ is divided into $200$ uniform cells. The Neumann boundary condition $\frac{\partial f}{\partial \bm n} =0$ is applied by extending the values on the boundary cells constantly along the outward-facing unit normal vector $\bm n$.
We test all three DVDM submodels with this problem.

The continuum regime is characterised with $\tau = 10^{-4}$. In all DVDM submodels, we set $N=8$ and the directions $\left \{ \bm l_m =  \left(\cos\frac{(2m-1)\pi}{16}, \sin\frac{(2m-1)\pi}{16} \right)^T \right \}_{m=1}^8$.
In the DVD-DVM, the discrete velocity nodes in each direction are selected as $\xi_k = 0.4 k - 5$ for $k=1,...,24$. In the DVD-HSM, we choose the order $M=12$ for the truncated series in Eq.~(\ref{eq:25_hsm_series}). The SSP2 scheme in Section~\ref{sec:schemes} is applied to both the DVD-DVM and DVD-HSM.
For the DVD-EQMOM, we set $M=2$.

\begin{figure}[htbp]
    \centering
    \includegraphics[height=7cm]{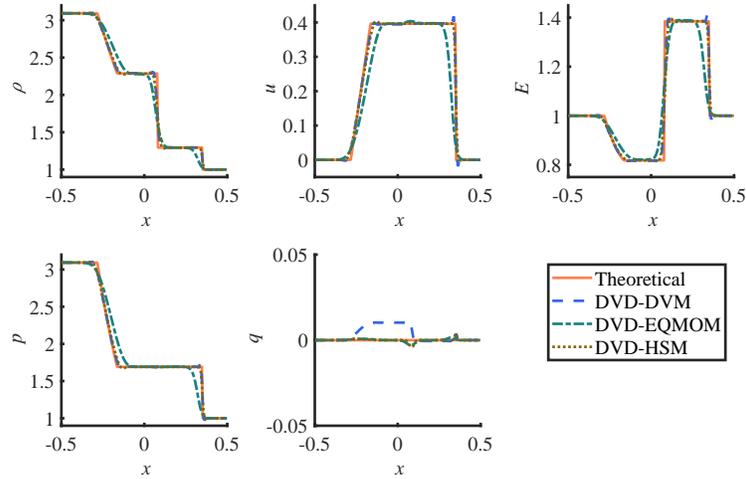}
    \caption{1-D Riemann problem with $\tau = 10^{-4}$: profiles of density $\rho$, velocity $u$, energy $E$, pressure $p$ and heat flux $q$ at $t = 0.2$. In all models, we set $N=8$ and the directions $\left\{ \bm l_m = \left(\cos\frac{(2m-1)\pi}{16},\sin\frac{(2m-1)\pi}{16} \right)^T \right \}_{m=1}^8$.
        In the DVD-DVM, the discrete velocities in each direction are $\xi_k = 0.4 k - 5$ for $k=1,...,24$.
        We set $M=12$ for the DVD-HSM and $M=2$ for the DVD-EQMOM.
    }\label{fig:1DRP_eu}
\end{figure}

Fig.~\ref{fig:1DRP_eu} shows the spatial distributions of the macroscopic quantities $(\rho, u, E, p)$ at $t=0.2$. Both the simulated results and theoretical solutions are plotted. The shock wave that goes right, the rarefraction wave that goes left, and the discontinuity between them are all well captured.
It is seen that what produced by both the DVD-DVM and DVD-HSM agree well with the analytical solutions except some oscillations near the discontinuities, while the two-node EQMOM is less accurate.
However, the DVD-DVM yields the worst result for the heat flux $\bm q = \frac{1}{2} \left \langle (\bm \xi - \bm U) (|\bm \xi - \bm U|^2 + |\bm \zeta|^2) f \right \rangle$, which should be zero since it is easy to verify that $\left \langle (\bm \xi - \bm U) (|\bm \xi - \bm U|^2 + |\bm \zeta|^2) \mathcal{E}[f] \right \rangle = 0$
for $\mathcal{E}[f]$ in Eq.~(\ref{eq:2_bgk_equil}) (the bracket $\langle\cdot\rangle$ is defined in Eq.~(\ref{eq:3_macpara})).
More directions and discrete nodes may be needed to reduce such a discrepancy.

\begin{figure}[htbp]
    \centering
    \includegraphics[height=7cm]{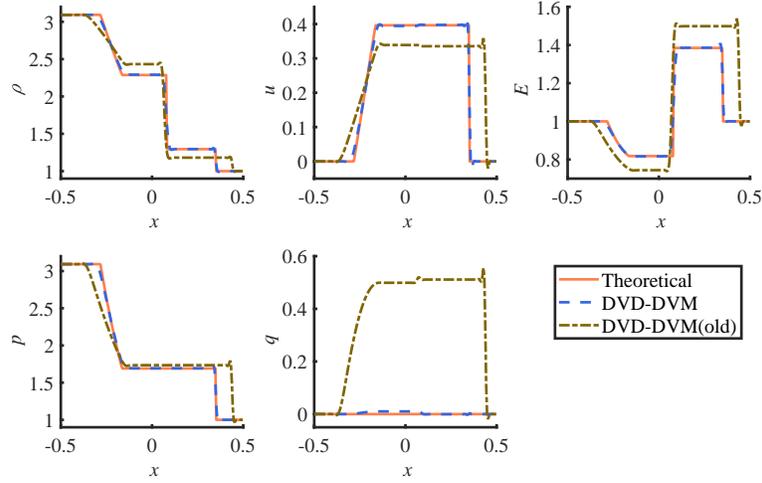}
    \caption{1-D Riemann problem with $\tau = 10^{-4}$: A comparison between the DVD-DVM predictions with and without the weight function $|\xi|^{D-1}$. All other setups are the same as in the previous case.
    }\label{fig:1DRP_compare_wrong}
\end{figure}

We emphasize that the weighted integral in Eq.~(\ref{eq:8_macpara_dvdm}), with the weight function $|\xi|^{D-1}$, is a key feature different from our previous model in \cite{Huang_2022_arxiv}. This weight function has been carefully treated in all DVDM submodels in Section~\ref{sec:submodels}.
As a direct comparison, Fig.~\ref{fig:1DRP_compare_wrong} shows that without this weight function, the predicted heat flux $q$ deviates significantly from zero, which contradicts the Euler limit solution. Other properties have larger errors as well.
Therefore, only with this weight function $|\xi|^{D-1}$, the resultant DVDM can produce satisfactory results.

\begin{figure}[htbp]
    \centering
    \includegraphics[height=7cm]{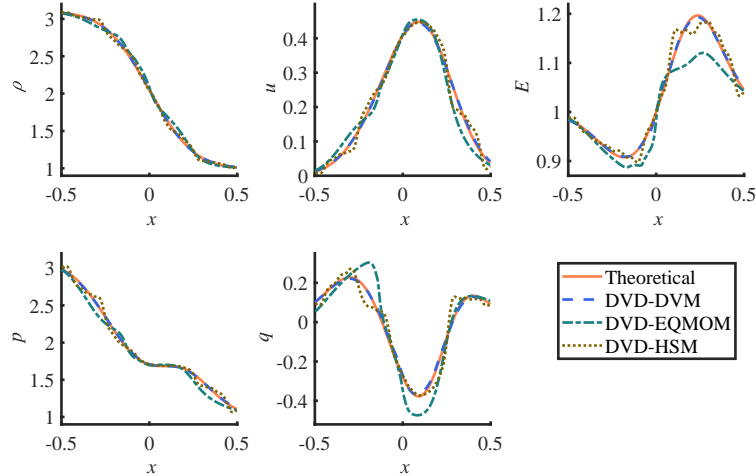}
    \caption{1-D Riemann problem with $\tau = 10^4$: profiles of density $\rho$, velocity $u$, energy $E$, pressure $p$ and heat flux $q$ at $t = 0.2$. In all models, we set $N=18$ and the directions $\left \{ \bm l_m =  \left(\cos\frac{(2m-1)\pi}{36},\sin\frac{(2m-1)\pi}{36} \right)^T \right\}_{m=1}^{18}$.
        In the DVD-DVM, the discrete velocities in each direction are $\xi_k = 0.4 k - 7.4$ for $k=1,...,36$.
        We set $M=12$ for the DVD-HSM and $M=2$ for the DVD-EQMOM.
    }\label{fig:1DRP_nc}
\end{figure}

As for the free-molecular flow regime, we take $\tau = 10^4$ to create a near-zero collision term. In all DVDM submodels, we set $N=18$ and the directions $\left \{ \bm l_m = \left (\cos\frac{(2m-1)\pi}{36},\sin\frac{(2m-1)\pi}{36} \right )^T \right \}_{m=1}^{18}$.
In the DVD-DVM, the discrete velocity nodes in each direction are selected as $\xi_k = 0.4 k - 7.4$ for $k=1,...,36$. In the DVD-HSM, we still choose the order $M=12$. The SSP2 scheme is applied to both the DVD-DVM and DVD-HSM. For the DVD-EQMOM, we set $M=2$.
Fig.~\ref{fig:1DRP_nc} presents the resulting profiles of macroscopic quantities at $t=0.2$. Obviously there is no shock in this case, and the DVD-DVM shows the highest accuracy. The relatively large error of the DVD-EQMOM is partly due to the small number of nodes ($M=2$) used in the simulation.

\subsection{Couette Flow}
\label{sec:Couette_flow}

The flow is confined between two infinite parallel walls located at $x = \pm 0.5 H$. The left and right walls move with constant velocities $\pm v_w \bm e_y$ to drive the fluid between them to a steady state. In this way, the flow reduces to a spatially 1-D problem in $x$. Assume $D=2$ and $L=0$ (no internal degrees of freedom).
Let $H = 1$, $v_w = 0.1$ and the wall temperature $\theta_w=2$.
The initial values of the fluid are $(\rho_0, \bm U_0, \theta_0) = (1,\bm 0, 2)$. These settings ensure a small Mach number.

In the Couette flow, different flow regimes are characterized by the parameter $\kappa := (\sqrt{\pi}/2) \mathrm{Kn}$, where the Knudsen number $\mathrm{Kn}$ is defined as \cite{Guo_2013}
\[
\mathrm{Kn} = \frac{\tau}{H} \sqrt{\frac{\pi \theta_0}{2}}.
\]
Thus, the flow regime can be tuned by varying the values of $\tau$.

Both the DVD-DVM and DVD-EQMOM are used with the first-order upwind scheme (see Section~\ref{sec:schemes}). The 1-D physical domain $[-0.5,0.5]$ is divided into $200$ uniform cells. The diffuse-scattering law is applied as the wall boundary condition. The computation stops when the $L^2$-norm of the difference of $\bm U$ between two consecutive time steps is smaller than $10^{-6}$, which indicates that the flow is in a steady state.
We set $N=15$ and the directions $\bm l_{\mathcal L} = \left \{ \left(\cos\frac{(m-1)\pi}{15},\sin\frac{(m-1)\pi}{15} \right)^T \right \}_{m=1}^{15}$ in all computations. For the DVD-DVM, the velocity nodes in each direction are chosen as $\xi_k = 0.5k -5.75$ for $k=1,...,22$. For the DVD-EQMOM, we let $M=2$.

\begin{figure}[htbp]
    \centering
    \subfigure{
        \includegraphics[height=4.5cm]{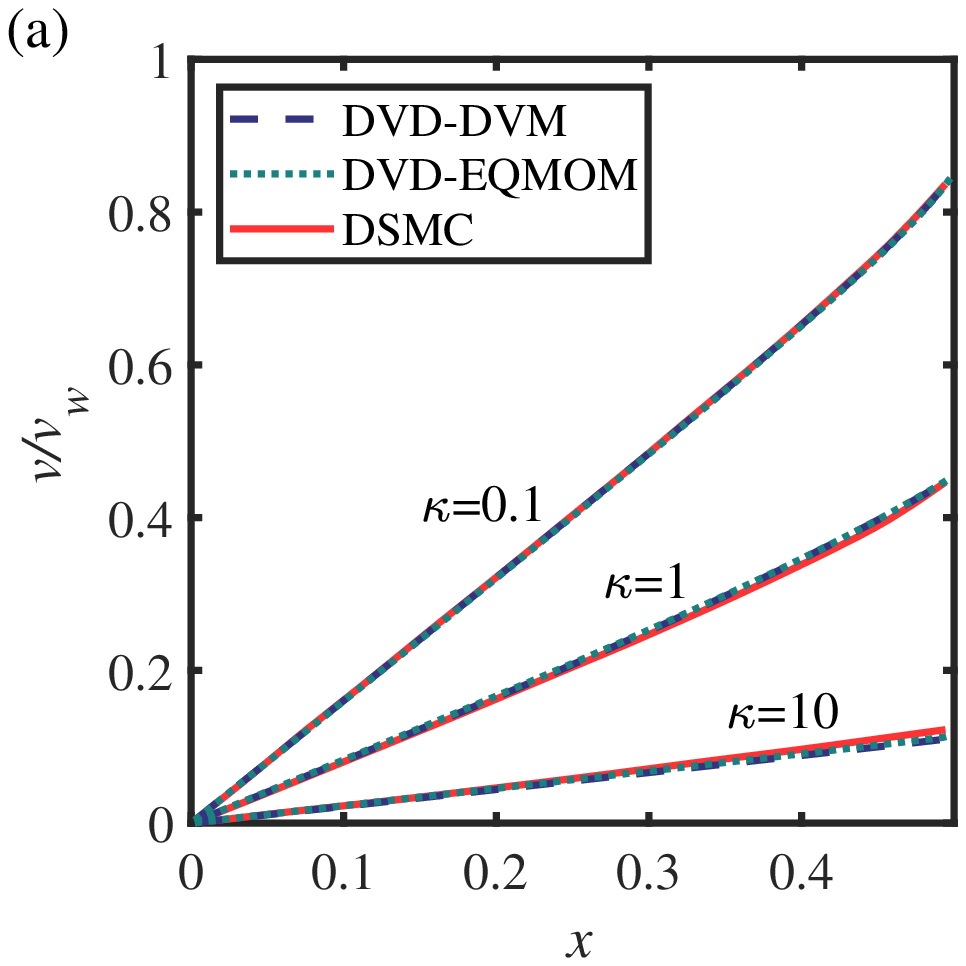}
        \label{fig:Couette_v}
    }%
    \subfigure{
        \includegraphics[height=4.5cm]{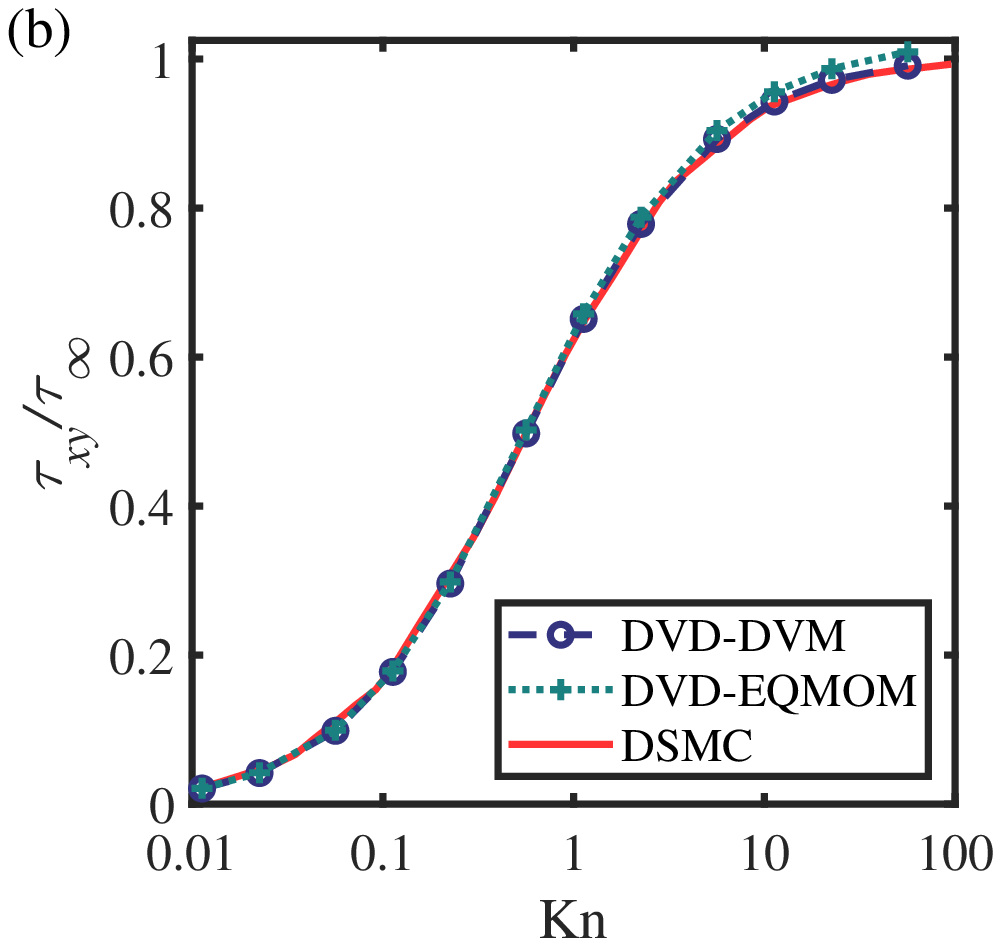}
        \label{fig:Couette_shear}
    }%
    \caption{Couette flow: (a) Steady-state vertical velocity profiles, and (b) Shear stress for different values of Kn. In all models, $N=15$ and the directions are $\{\bm l_m = (\cos\frac{(m-1)\pi}{15},\sin\frac{(m-1)\pi}{15})\}_{m=1}^{15}$. In the DVD-DVM, the discrete velocities in each direction are $\xi_k = 0.5k -5.75$ for $k=1,...,22$. The DSMC data is from \cite{Bahukudumbi_2003}.
    }\label{fig:Couette}
\end{figure}

Fig.~\ref{fig:Couette_v} shows the steady-state vertical velocity profiles on the positive domain $x>0$ for different values of $\kappa$. The velocity is normalized by the wall velocity $v_w$. The DSMC results in \cite{Bahukudumbi_2003} are included for a comparison. Apparently, higher values of $\kappa$ correspond to more rarefied gases and less momentum transfer from the moving wall to the fluids. Both the DVD-DVM and DVD-EQMOM reproduce the velocity profiles quite close to the reference data for all three values of $\kappa$.
Fig.~\ref{fig:Couette_shear} further presents the shear stress $\tau_{xy}$ defined by
\[
\tau_{xy} = \int_{\mathbb{R}^2} (\xi_x - u)(\xi_y - v) f(\bm \xi) d \bm \xi = \int_{\mathbb{R}^2} \xi_x \xi_y f(\bm \xi) d \bm \xi - \rho u v
\]
for Kn ranging from 0.01 to 100. Here we denote $\bm \xi = (\xi_x,\xi_y)^T$ and $\bm U=(u,v)^T \in \mathbb R^2$. The shear stress is normalized by the free-molecular stress $\tau_{\infty} = -\rho u_w \sqrt{2 \theta / \pi}$.
Our DVDM results are generally in good agreement with the DSMC results \cite{Bahukudumbi_2003}. It is seen that the two-node DVD-DVDM has more significant errors at larger Kn (rarefied flow) conditions, as compared with the DVD-DVM made up by more velocity nodes.

\subsection{2-D Riemann Problems}
\label{sec:2D_Riemann_prob}

Two-dimensional Riemann problems have been studied in \cite{Lax_1998}. Here we consider the following initial data
\[
(\rho,u,v,p) =
\left\{
\begin{array}{lll}
    (\rho_1,u_1,v_1,p_1)=(0.5313,\ 0,\ 0,\ 0.4), & x>0, & y>0,\\
    (\rho_2,u_2,v_2,p_2)=(1,\ 0.7276,\ 0,\ 1), & x\leq 0, & y>0,\\
    (\rho_3,u_3,v_3,p_3)=(0.8,\ 0,\ 0,\ 1), & x\leq 0, & y\leq 0,\\
    (\rho_4,u_4,v_4,p_4)=(1,\ 0,\ 0.7276,\ 1), & x>0, & y\leq 0,
\end{array}
\right.
\]
which was also studied in \cite{Guo_2015}. In contrast to the previous subsections, the internal degrees of freedom is involved here. Thus we set $L=3$ and the specific heat ratio $\gamma=(2+D+L)/(D+L)=1.4$.
The computational domain is $[-0.5,0.5]^2$. The Neumann condition is applied on the boundary. Like in Subsection \ref{sec:1D_Riemann_prob}, only the continuum and collisionless limits are considered.

\begin{figure}[htbp]
    \centering
    \subfigure{
        \includegraphics[height=5cm]{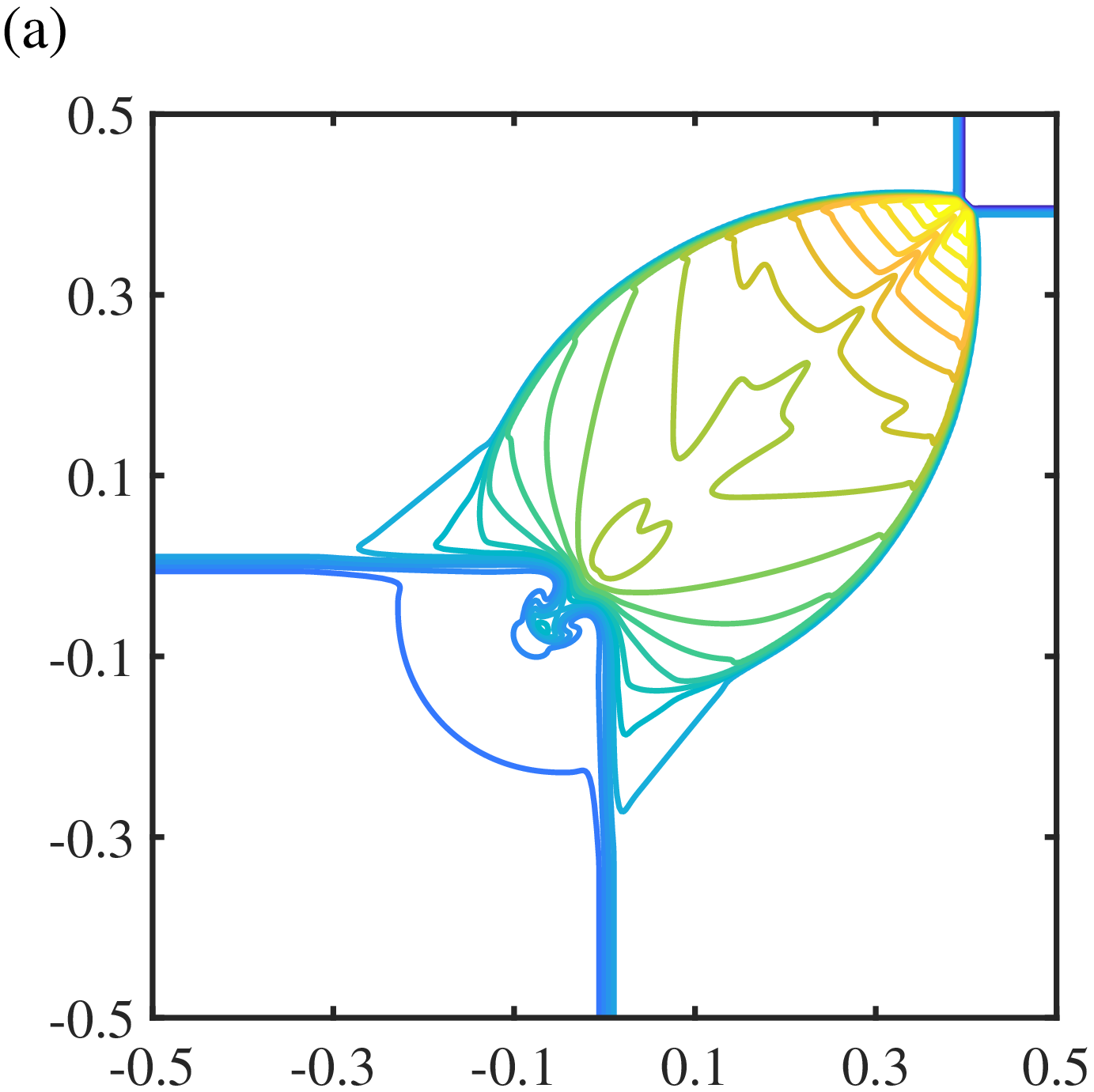}
        \label{fig:2DRP_eu_DVM}
    }%
    \subfigure{
        \includegraphics[height=5cm]{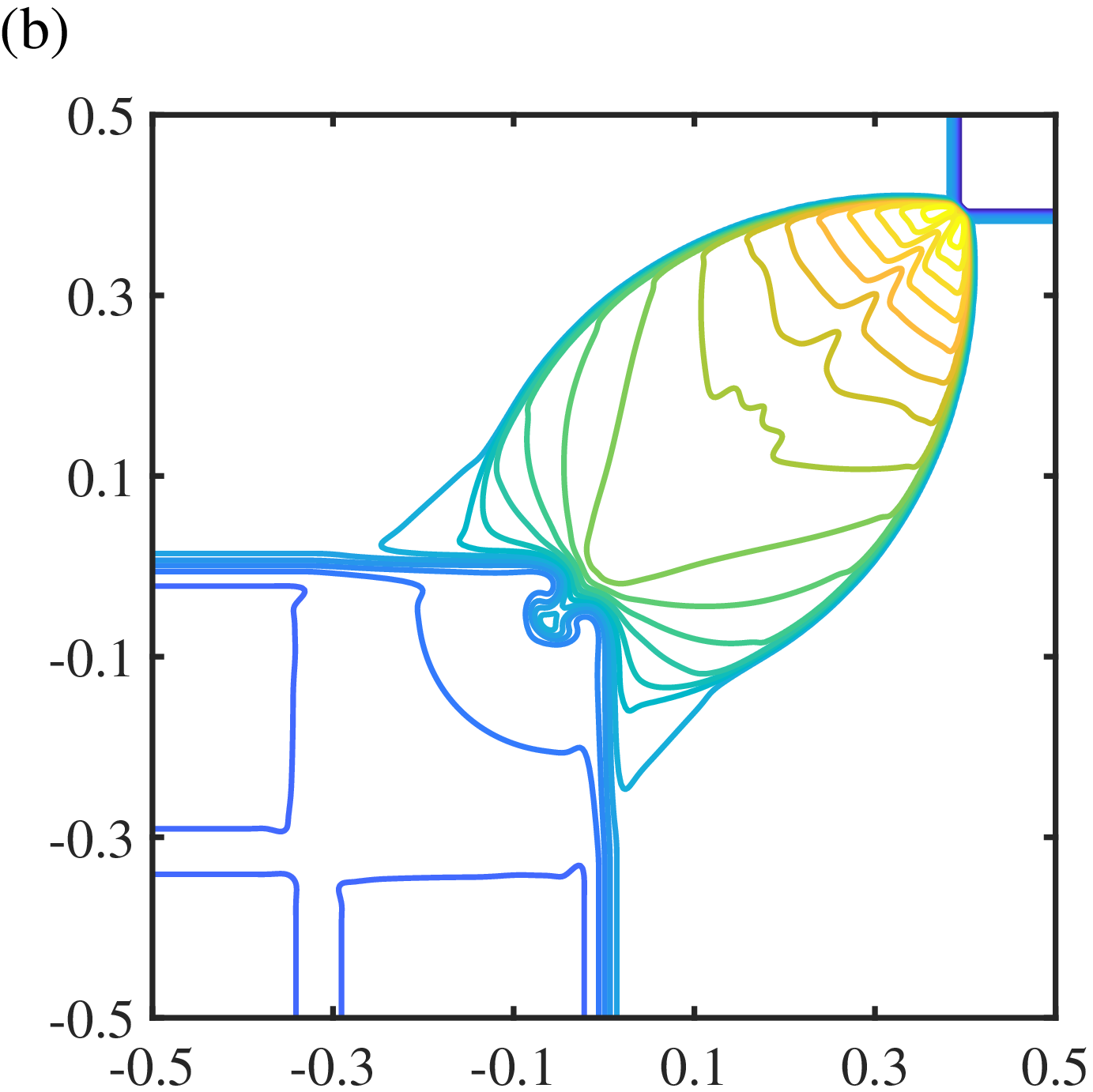}
        \label{fig:2DRP_eu_HSM}
    }%
    \caption{2-D Riemann problem with $\tau=10^{-4}$: density contours at $t = 0.25$ simulated by (a) DVD-DVM and (b) DVD-HSM.
        Let $N=8$ and $\bm l_m = \left(\cos \frac{(2m-1)\pi}{16}, \sin \frac{(2m-1)\pi}{16} \right)^T$ for $m=1,...,8$ in both models. For the DVD-DVM, the discrete velocity nodes in each direction are taken as $\xi_k = k -8.5$ for $k=1,...,16$. For the DVD-HSM, we have $M=12$.
    } \label{fig:2D_Riemann_eu}
\end{figure}

\begin{figure}[htbp]
    \centering
    \subfigure{
        \includegraphics[height=5cm]{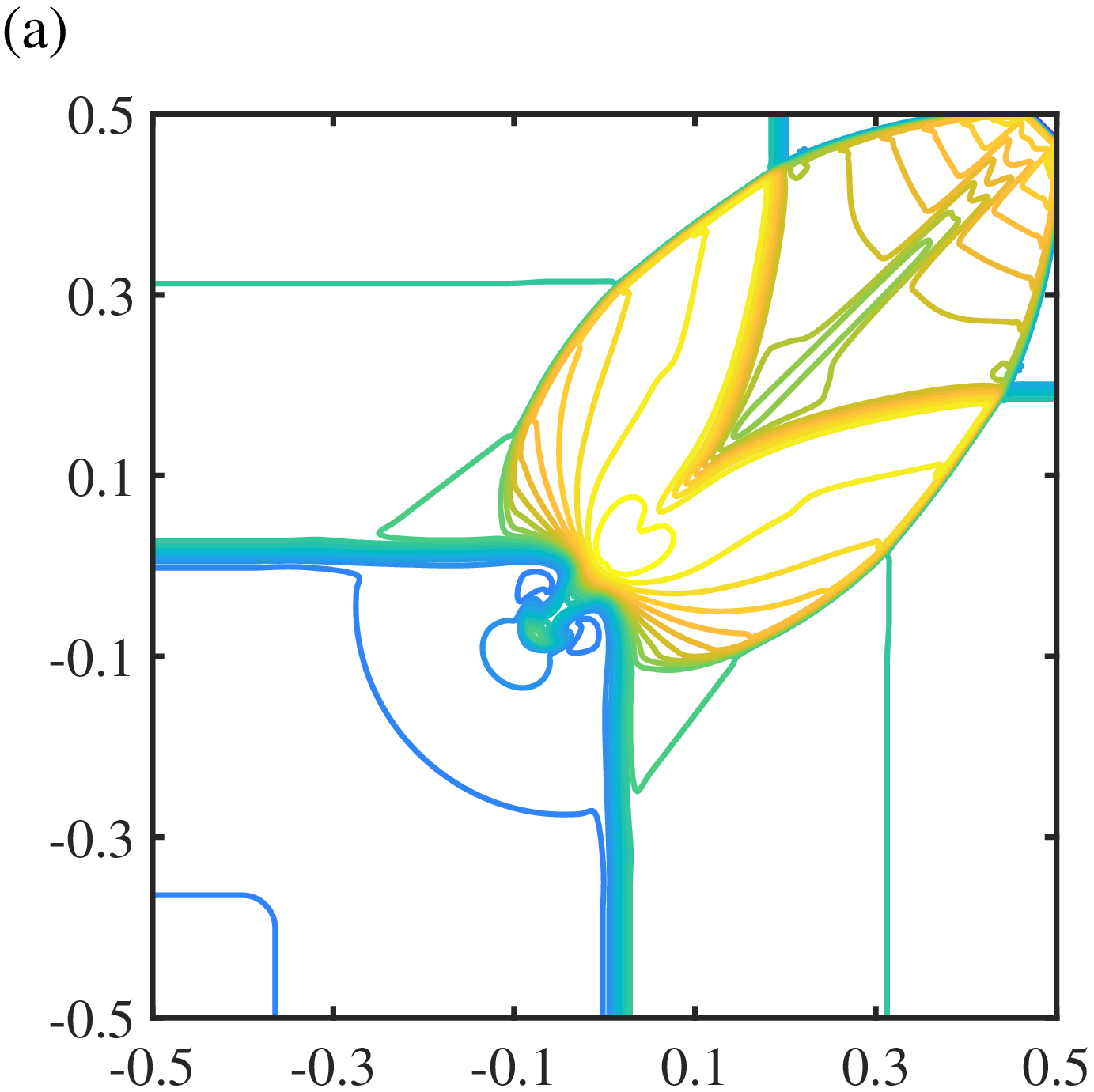}
        \label{fig:2DRP_eu_wrong_DVM}
    }%
    \subfigure{
        \includegraphics[height=5cm]{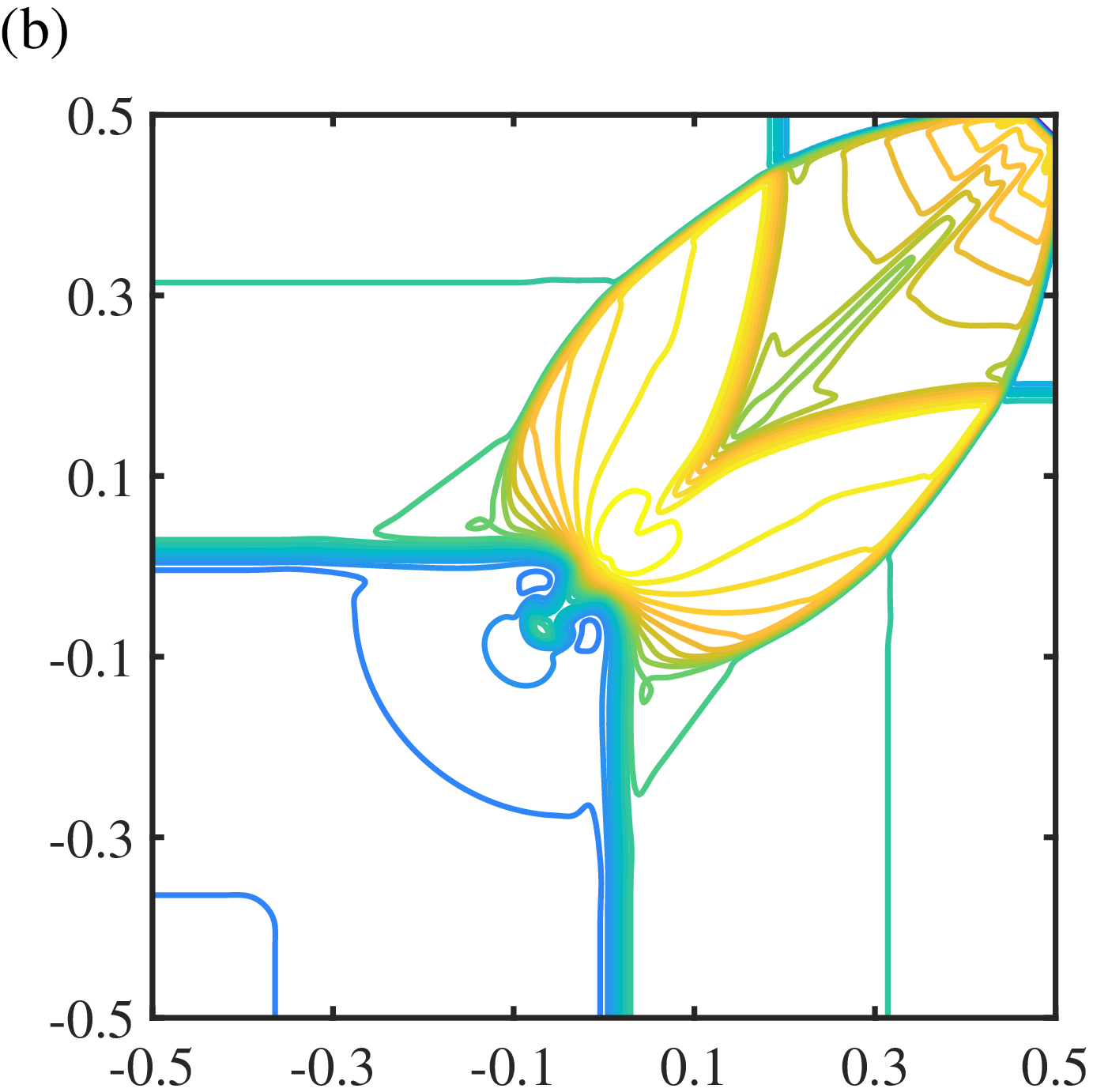}
        \label{fig:2DRP_eu_wrong_HSM}
    }%
    \caption{2-D Riemann problem with $\tau=10^{-4}$: density contours at $t = 0.25$ simulated by (a) DVD-DVM and (b) DVD-HSM without the weight function $|\bm \xi|^{D-1}$. All other setups are the same as in the previous case.
    }\label{fig:2D_Riemann_eu_wrong}
\end{figure}

The continuum regime is characterised again by $\tau = 10^{-4}$.
Both the DVD-DVM and DVD-HSM are tested in this case. We set $N=8$ and the directions $\bm l_{\mathcal L} =\left\{ \left(\cos \frac{(2m-1)\pi}{16}, \sin \frac{(2m-1)\pi}{16} \right)^T \right \}_{m=1}^8$.
For the DVD-DVM, the discrete velocity nodes in each direction are taken as $\xi_k = k -8.5$ for $k=1,...,16$. For the DVD-HSM, we set $M=12$. The SSP2 scheme is applied for both models.
The physical domain $[-0.5,0.5]^2$ is divided into a $400\times 400$ uniform mesh. Fig.~\ref{fig:2D_Riemann_eu} shows the density contours at $t = 0.25$ simulated by the both models. The shock waves and contact discontinuities are clearly manifested, which agree reasonably well with the solutions of kinetic equation in \cite{Guo_2015} and Euler equation in \cite{Lax_1998}.
We again remark that the weight $|\bm \xi|^{D-1}$ in the DVDM Eq.~(\ref{eq:8_macpara_dvdm}) is necessary. As is revealed in Fig.~\ref{fig:2D_Riemann_eu_wrong}, if such a weight is absent, neither the DVD-DVM nor DVD-HSM correctly predicts the density contour for the Riemann problem in the continuum limit.

The collisionless free-molecular regime is characterised with $\tau = 10^4$.  The analytical results can be found in \cite{Guo_2015}.
All three DVDM submodels are used in the simulation. The non-equilibrium flow generally requires more elaborated discretization of the velocity space than the continuum case, while the absence of shock or discontinuity allows greater sizes of the spatial cells.
Thus, the physical domain $[-0.5,0.5]^2$ is discretized into a $80\times 80$ uniform mesh.
For the DVD-DVM, we set $N=24$ and the directions $\bm l_m = \left( \cos \frac{(2m-1)\pi}{48}, \sin \frac{(2m-1)\pi}{48} \right)^T$. The discrete velocity nodes in each direction are taken as $\xi_k = 0.4k - 9.8$ for $k=1,...,48$.
Remark that the total number of velocity nodes 1152 is much smaller than that used in \cite{Guo_2015} (over 40 000).
For the DVD-HSM, we set $N=30$ and $M=14$. The SSP2 scheme is used for both the DVD-DVM and DVD-HSM.
For the DVD-EQMOM, we set $N=30$ and $M=2$. The upwind scheme is employed.

\begin{figure}[htbp]
    \centering
    \subfigure{
        \includegraphics[height=4.25cm]{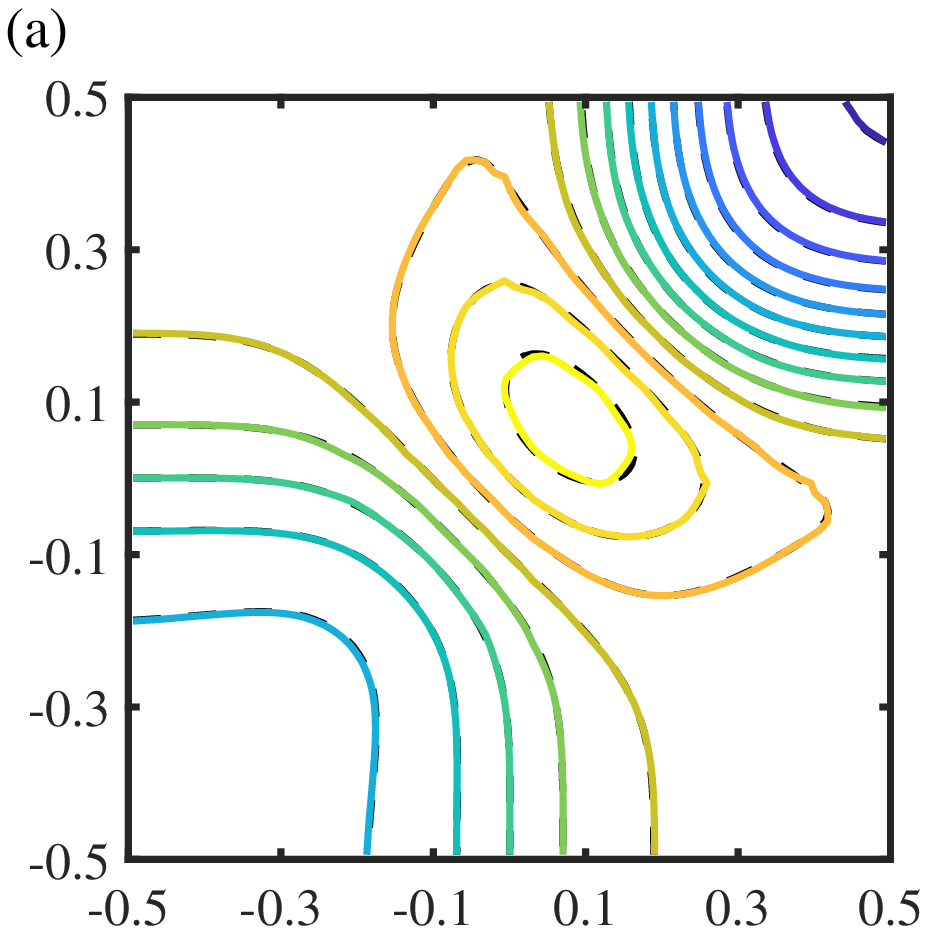}
        \label{fig:2DRP_nc_DVM_rho}
    }%
    \subfigure{
        \includegraphics[height=4.25cm]{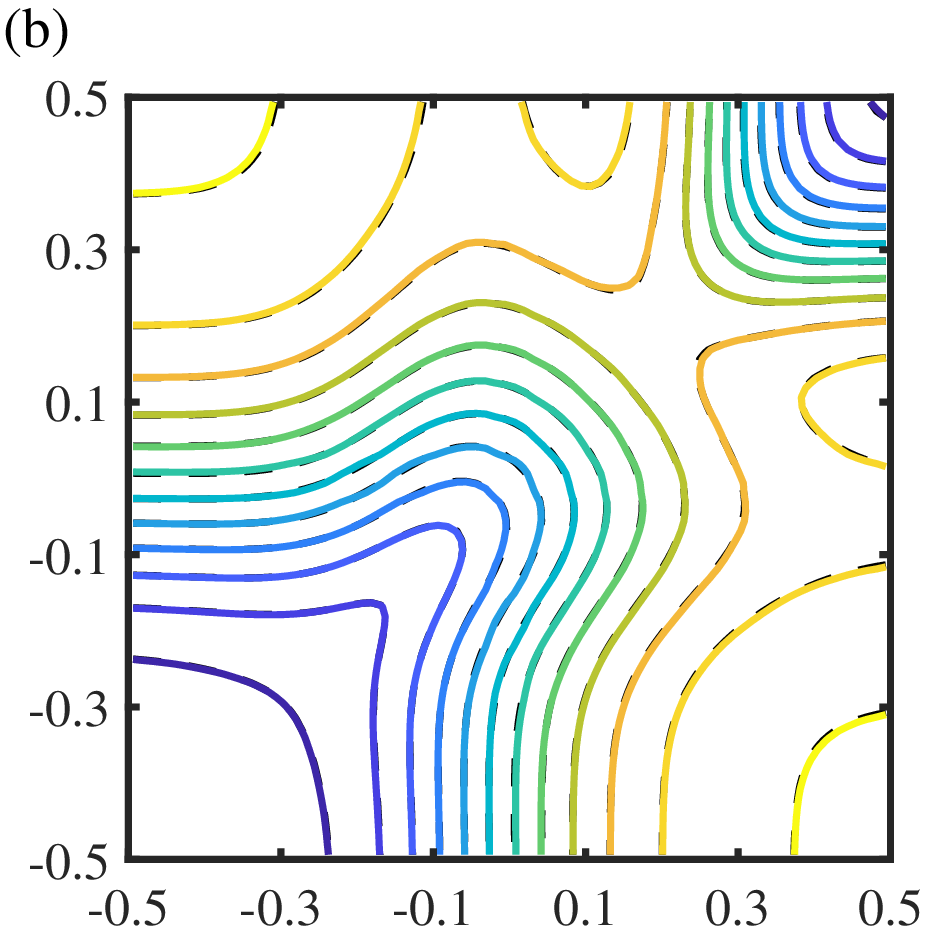}
        \label{fig:2DRP_nc_DVM_vm}
    }%
    \subfigure{
        \includegraphics[height=4.25cm]{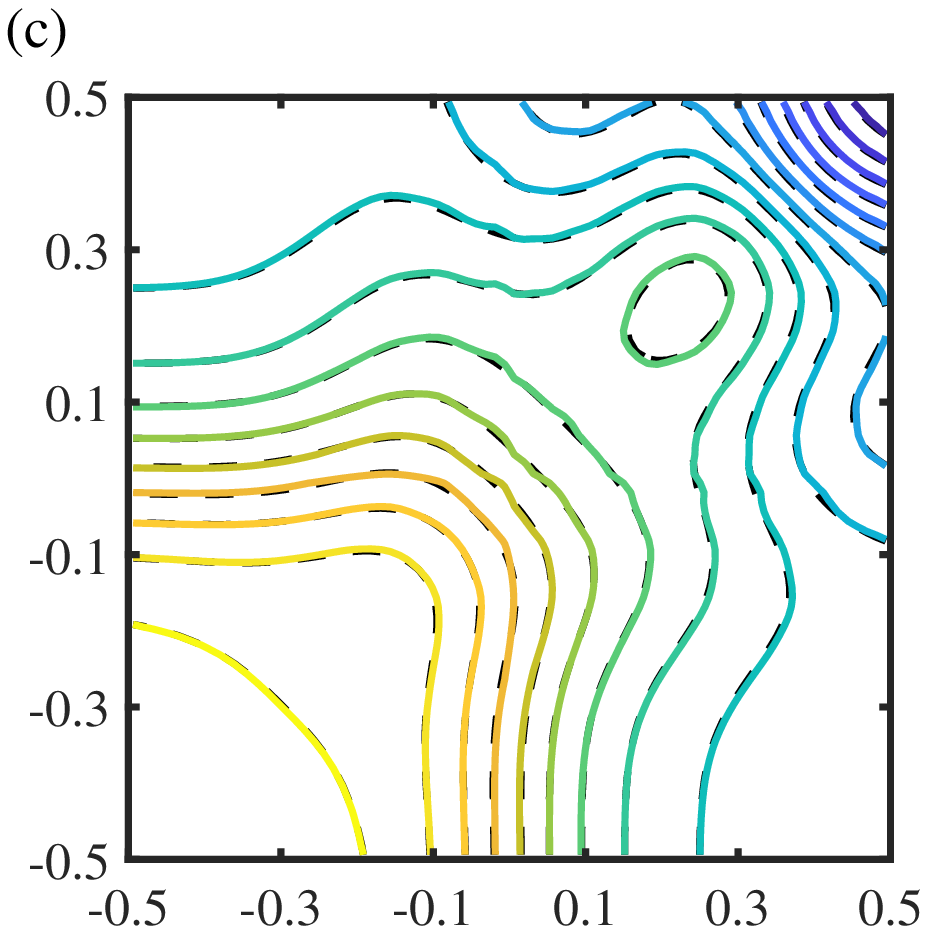}
        \label{fig:2DRP_nc_DVM_temp}
    }%
    \caption{2-D Riemann problem with $\tau=10^4$ by the DVD-DVM: contours of (a) density, (b) velocity magnitude and (c) temperature at $t = 0.15$. The analytical solutions are shown as black dashed lines. We set $N=24$ and the directions are of the similar form as before. Discrete velocities are $\xi_k = 0.4k - 9.8$ for $k=1,...,48$.
    }\label{fig:2D_Riemann_nc_DVM}
\end{figure}

\begin{figure}[htbp]
    \centering
    \subfigure{
        \includegraphics[height=4.25cm]{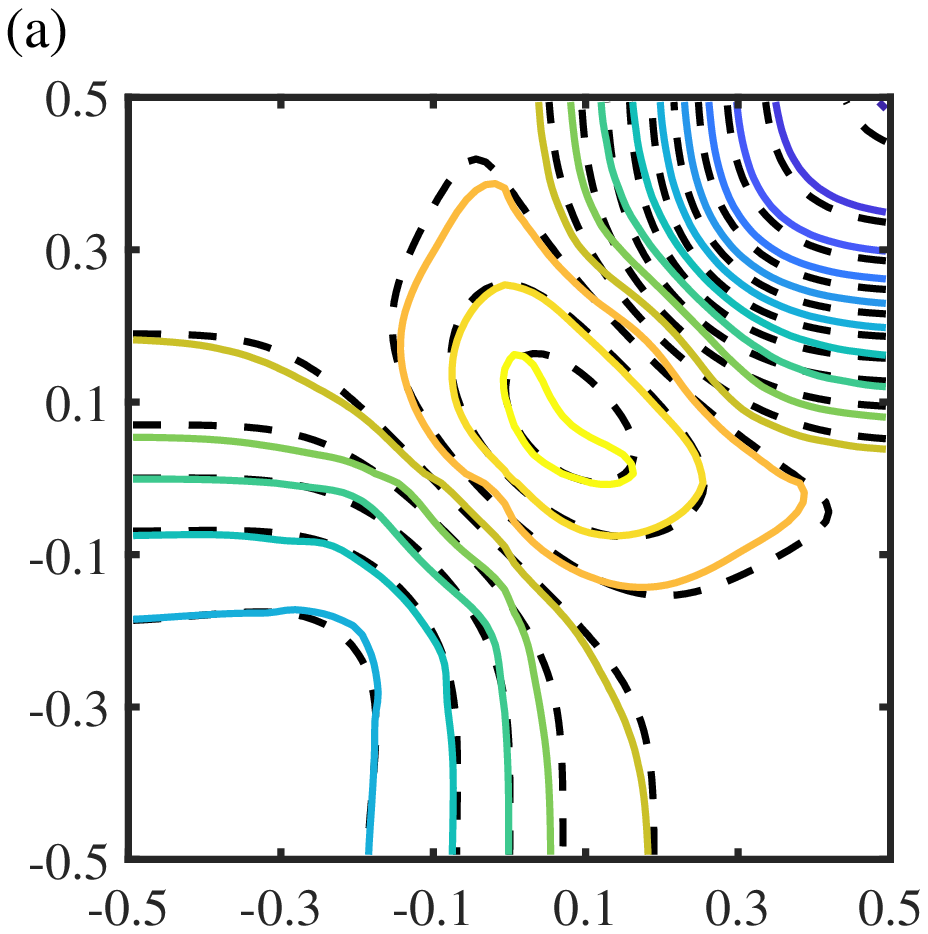}
        \label{fig:2DRP_nc_EQMOM_rho}
    }%
    \subfigure{
        \includegraphics[height=4.25cm]{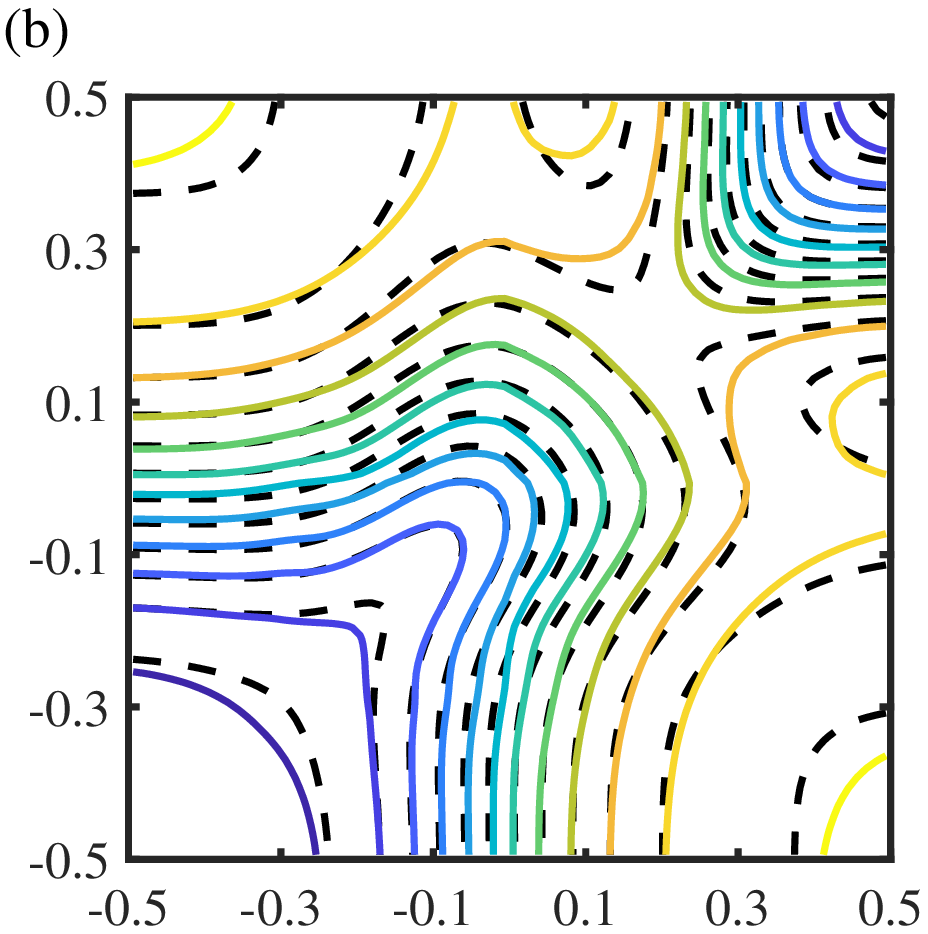}
        \label{fig:2DRP_nc_EQMOM_vm}
    }%
    \subfigure{
        \includegraphics[height=4.25cm]{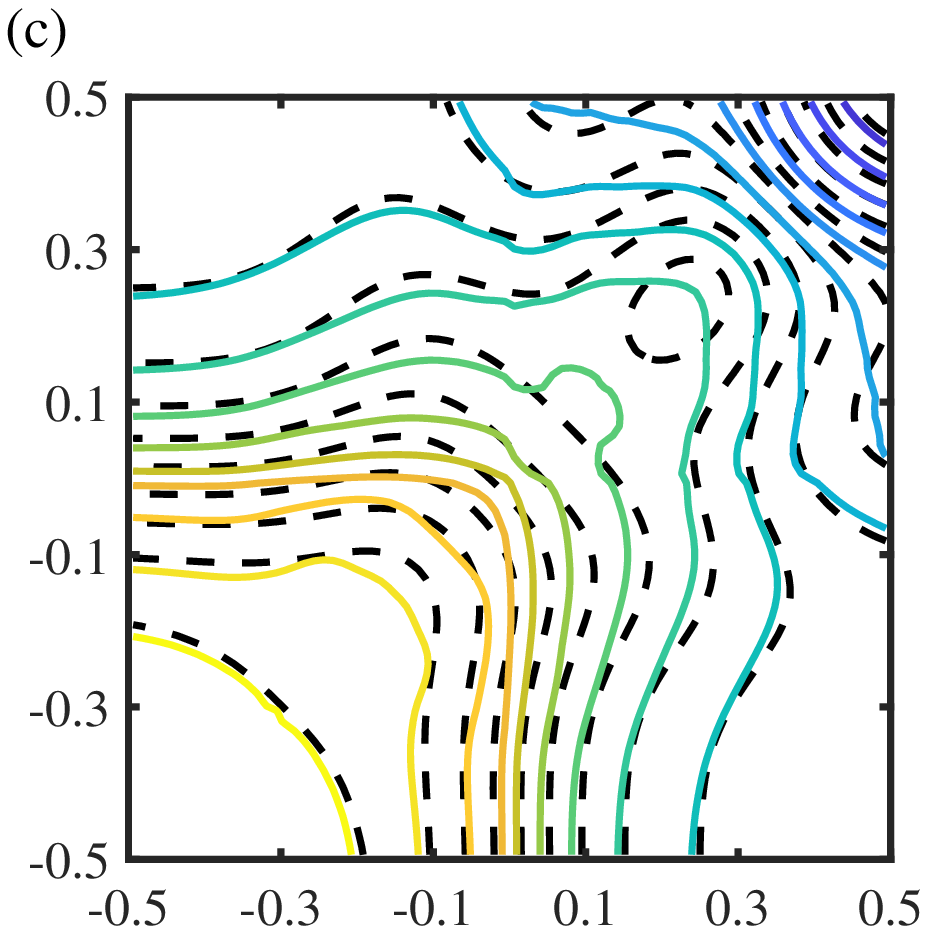}
        \label{fig:2DRP_nc_EQMOM_temp}
    }%
    \caption{2-D Riemann problem with $\tau=10^4$ by the DVD-EQMOM: contours of (a) density, (b) velocity magnitude and (c) temperature at $t = 0.15$. The analytical solutions are shown as black dashed lines. $N$ is set to be $30$.
    }\label{fig:2D_Riemann_nc_EQMOM}
\end{figure}

\begin{figure}[htbp]
    \centering
    \subfigure{
        \includegraphics[height=4.25cm]{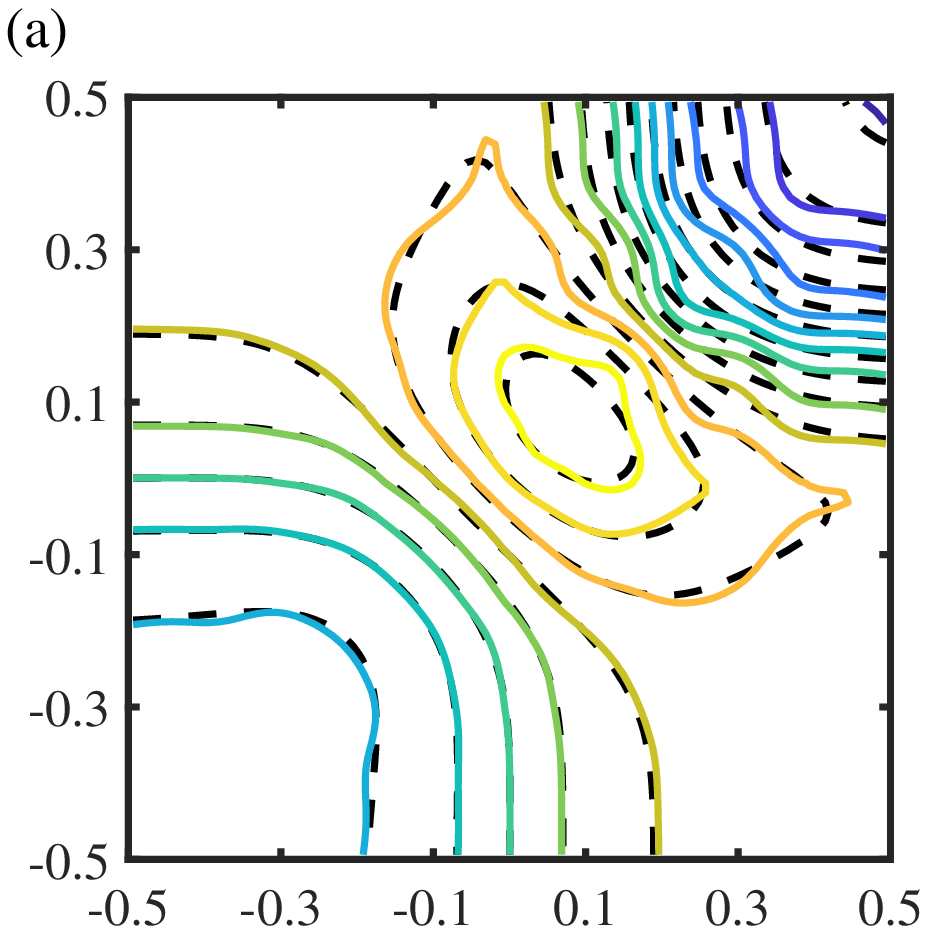}
        \label{fig:2DRP_nc_HSM_rho}
    }%
    \subfigure{
        \includegraphics[height=4.25cm]{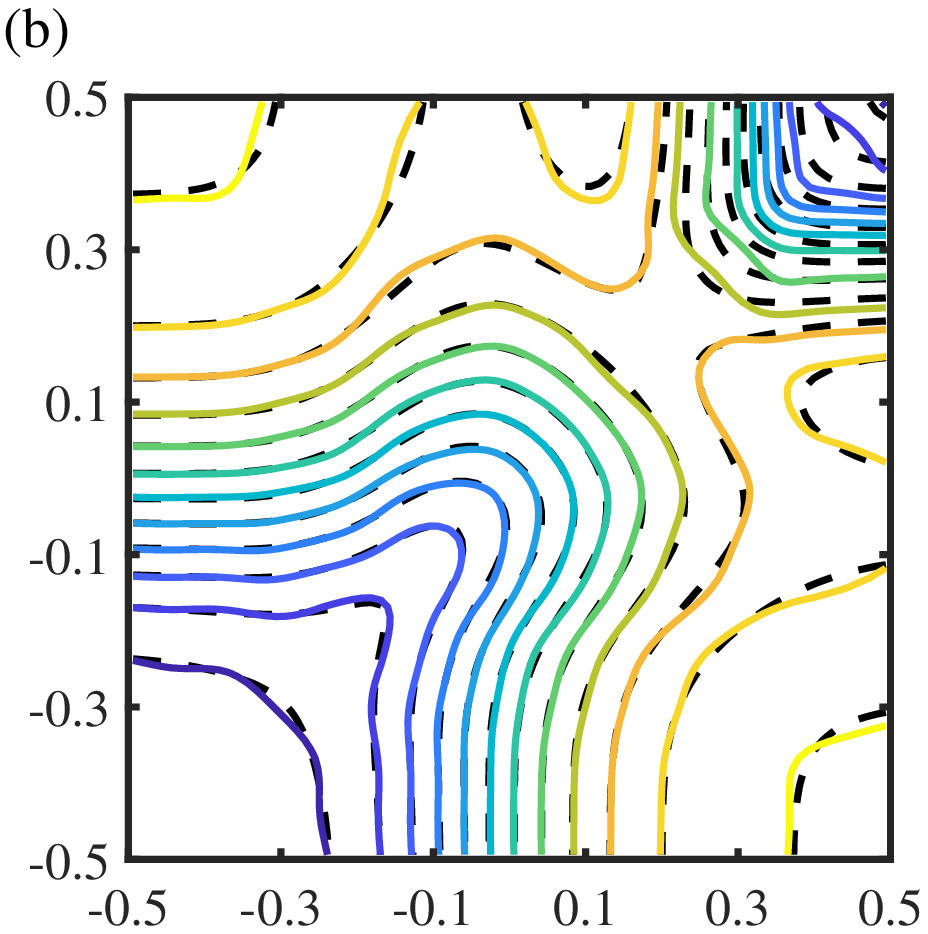}
        \label{fig:2DRP_nc_HSM_vm}
    }%
    \subfigure{
        \includegraphics[height=4.25cm]{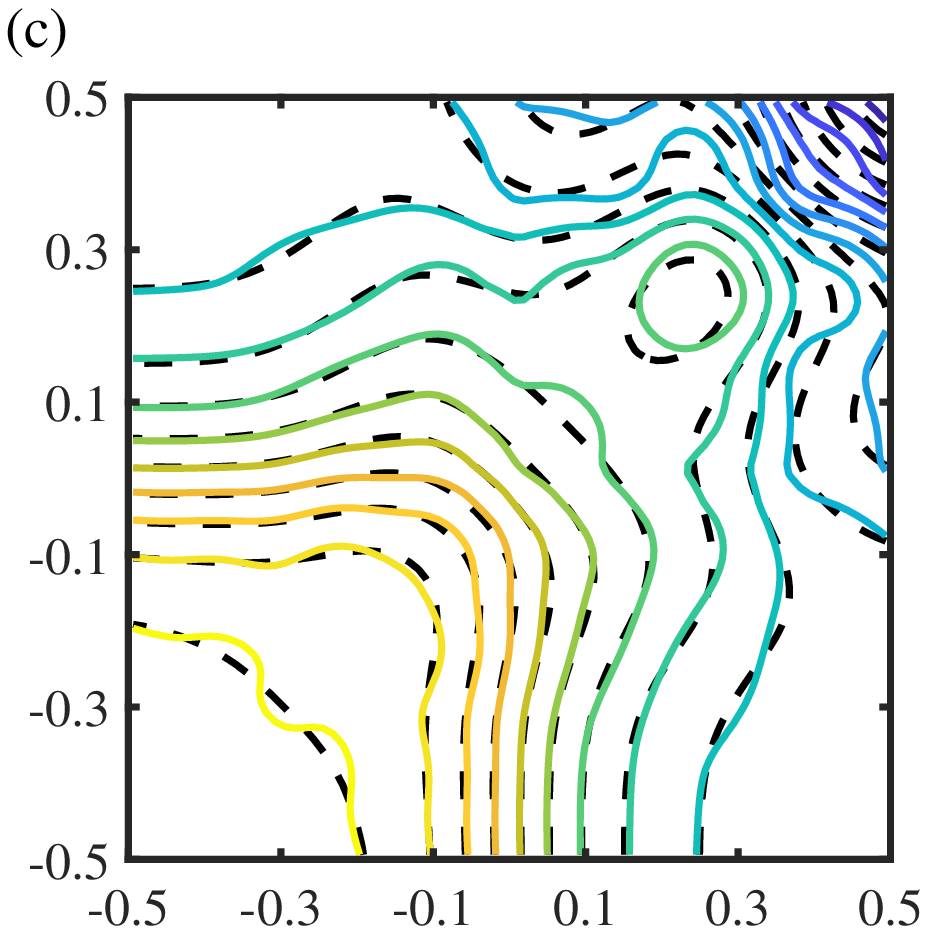}
        \label{fig:2DRP_nc_HSM_temp}
    }%
    \caption{2-D Riemann problem with $\tau=10^4$ by the DVD-HSM: contours of (a) density, (b) velocity magnitude and (c) temperature at $t = 0.15$. The analytical solutions are shown as black dashed lines. We set $N=30$ and $M=14$.
    }\label{fig:2D_Riemann_nc_HSM}
\end{figure}

The contours of density, temperature and velocity magnitude at $t=0.15$ are presented in Figs.~\ref{fig:2D_Riemann_nc_DVM}-\ref{fig:2D_Riemann_nc_HSM} by using different models. Also plotted are the analytical solutions (black dashed line). It is clearly seen that the DVD-DVM yields accurate predictions.
In contrast, the DVD-EQMOM and DVD-HSM exhibit greater errors, especially in the temperature profiles. This may be partly attributed to the lower-order approximation in the velocity space (i.e., small values of $M$) or lower-order discretization scheme (i.e., the first-order upwind scheme for the DVD-EQMOM).
Future work is needed to address these issues.

\subsection{Lid-Driven Cavity Flow}
\label{sec:lid_driven_flow}

Our last case is the two-dimensional lid-driven flow in a square cavity $[0,H]^2$. The upper wall moves horizontally with a constant speed $u_w$ to drive the fluid while the other three walls are fixed.
There are two types of lid-driven cavity flows. The first type is also termed as the microcavity flow, where the Reynolds number $\rm Re$ is so small that the flow is mainly characterized by the Knudsen number \cite{Naris2005,Guo_2013,John_2010}.
The other type with ${\rm Re} \gg 1$ has been widely studied by either solving the Navier-Stokes equation \cite{Ghia_1982} or employing the lattice Boltzmann method \cite{Hou1995}.

In this case, the internal degrees of freedom are neglected, that is, $D=2$ and $L=1$.
Our aim is to derive the steady-state flow field and the simulations start from a static flow ($\bm U = \bm 0$) in equilibrium with a constant density $\rho=1$ at $t=0$.
Let the initial temperatures of both the fluid and the walls be $\theta_0$ and assume that the walls keep this temperature. Then the upper wall starts to move and drive the fluid in the cavity.
The computation lasts until the flow becomes steady when the $L^2$-norm of the difference of $\bm U$ between two consecutive time steps is smaller than $10^{-6}$.

We first consider the microcavity flows where the Knudsen number ${\rm Kn} = \frac{\tau}{H} \sqrt{\frac{\pi \theta_0}{2}}$.
In this case, we set $H=1$, $\theta_0 = 2.4$ and $u_w = 0.32$, resulting in a Mach number of 0.16.
We thus tune $\rm Kn$ by taking different values of $\tau$.
Only the DVD-DVM is used to simulate the flow, for which we set $N=30$ and the directions $\bm l_m = \left(\cos\frac{(2m-1)\pi}{60}, \sin\frac{(2m-1)\pi}{60} \right)^T$ ($m=1,\dots,30$).
The discrete velocity nodes are taken as $\xi_k = 0.3k -8.85$ for $k=1,...,60$.
The cavity $[0,H]^2$ is divided into $100\times 100$ uniform cells. The upwind scheme incorporated with diffuse-scattering boundary laws is applied here.

\begin{figure}[htbp]
    \centering
    \subfigure{
        \includegraphics[height=4.25cm]{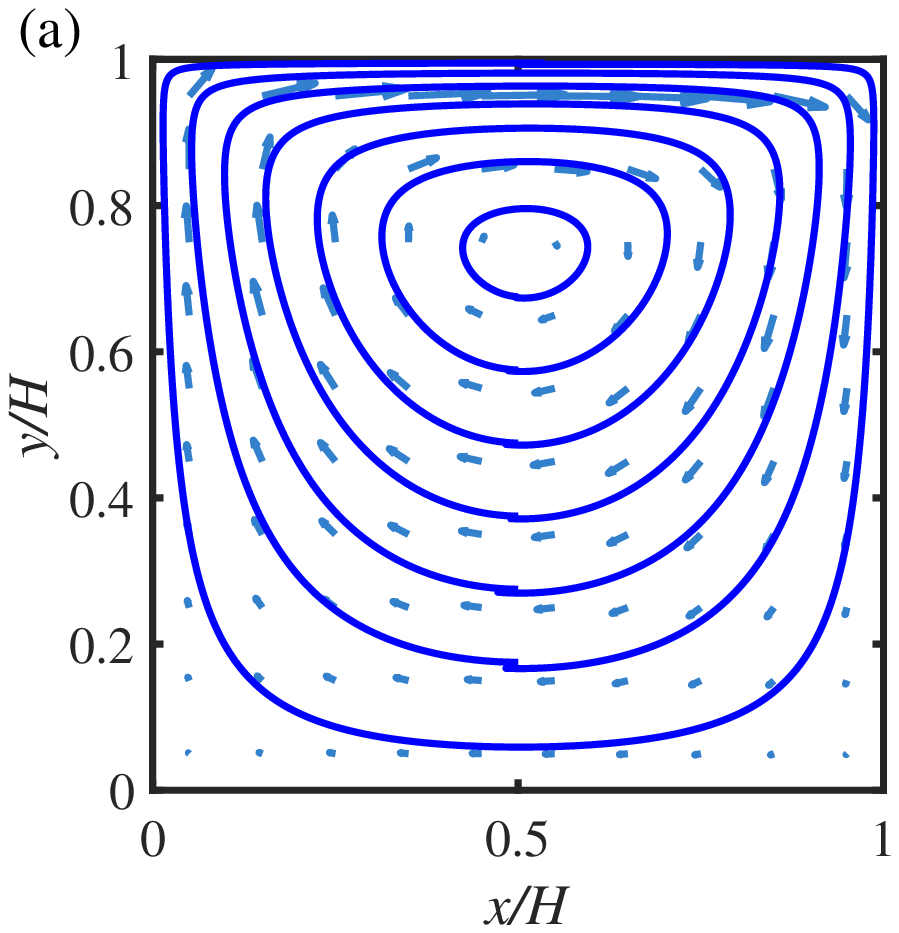}
        \label{fig:lid_driven_streamlines_Kn0_1}
    }%
    \subfigure{
        \includegraphics[height=4.25cm]{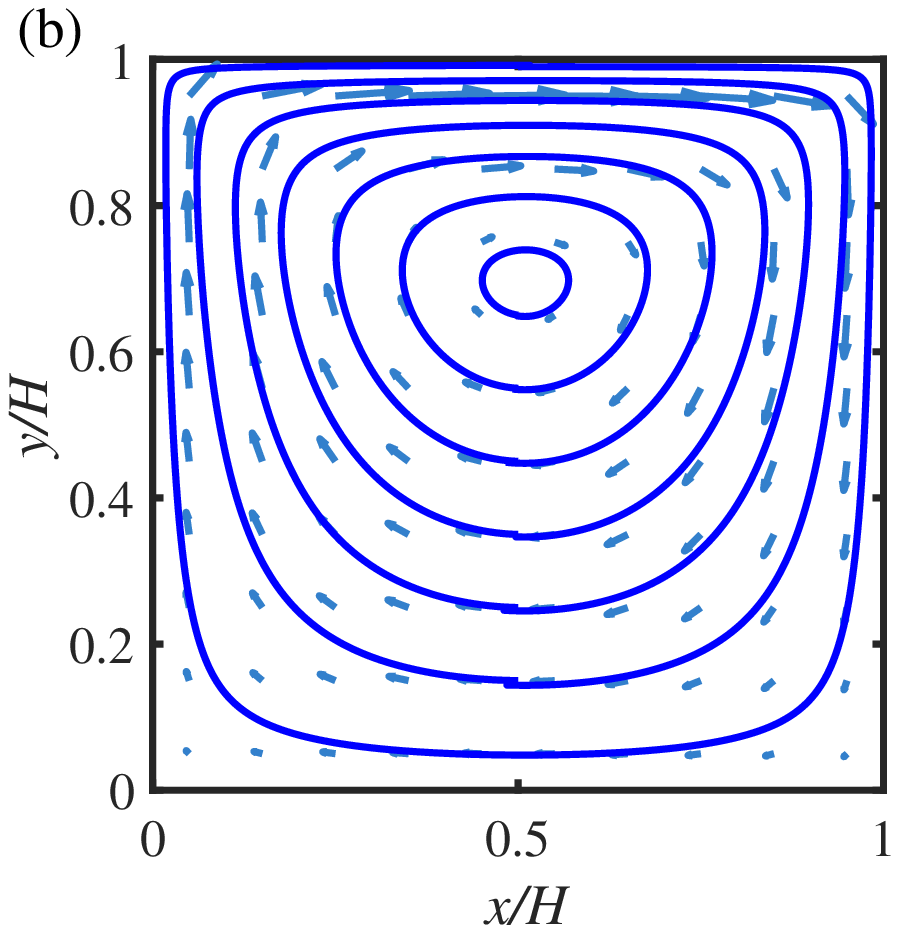}
        \label{fig:lid_driven_streamlines_Kn1}
    }%
    \subfigure{
        \includegraphics[height=4.25cm]{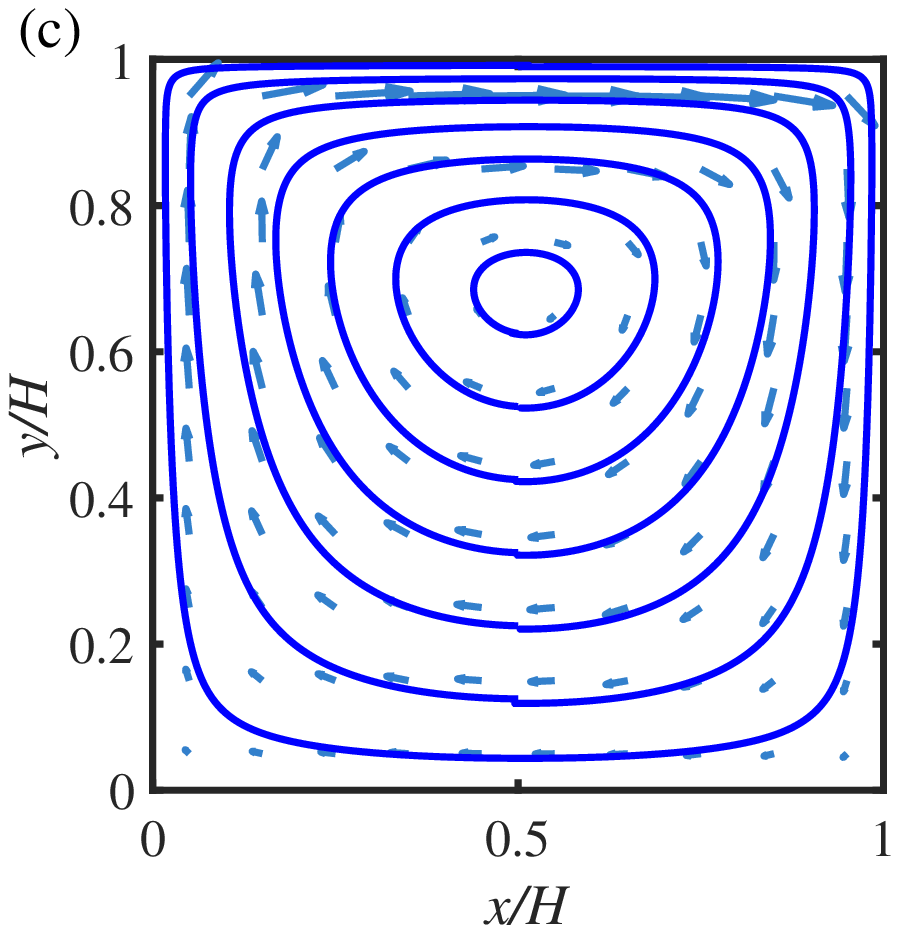}
        \label{fig:lid_driven_streamlines_Kn8}
    }%
    \caption{Microcavity flow: velocity streamlines of (a) Kn = 0.1, (b) Kn = 1 and (c) Kn = 8. For discrete directions and velocities, we set $N=30$ and $\{\bm l_m = (\cos\frac{(2m-1)\pi}{60}, \sin\frac{(2m-1)\pi}{60})\}_{m=1}^{30}$ while $\xi_k = 0.3k -8.85$ for $k=1,...,60$.
    }\label{fig:lid_driven_streamlines}
\end{figure}

\begin{figure}[htbp]
    \centering
    \subfigure{
        \includegraphics[height=4.5cm]{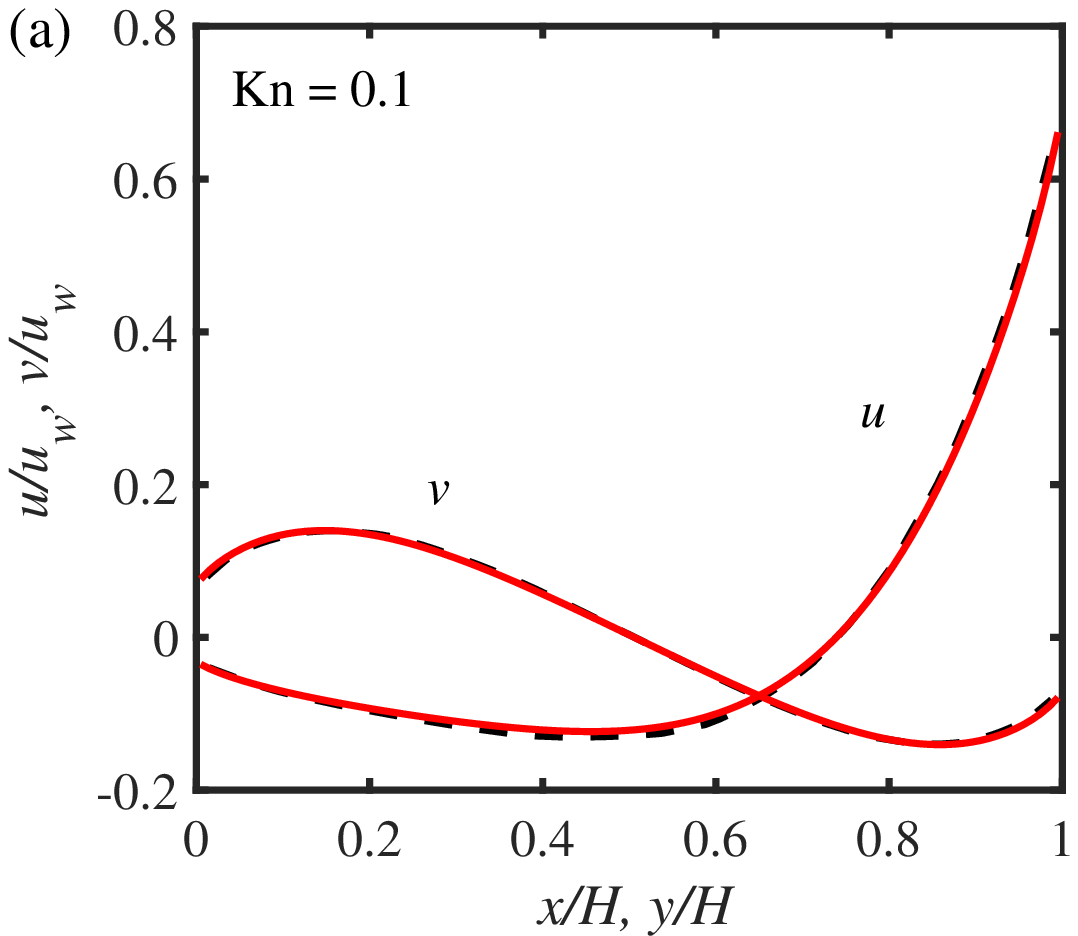}
        \label{fig:lid_driven_uv_Kn0_1}
    }%
    \subfigure{
        \includegraphics[height=4.5cm]{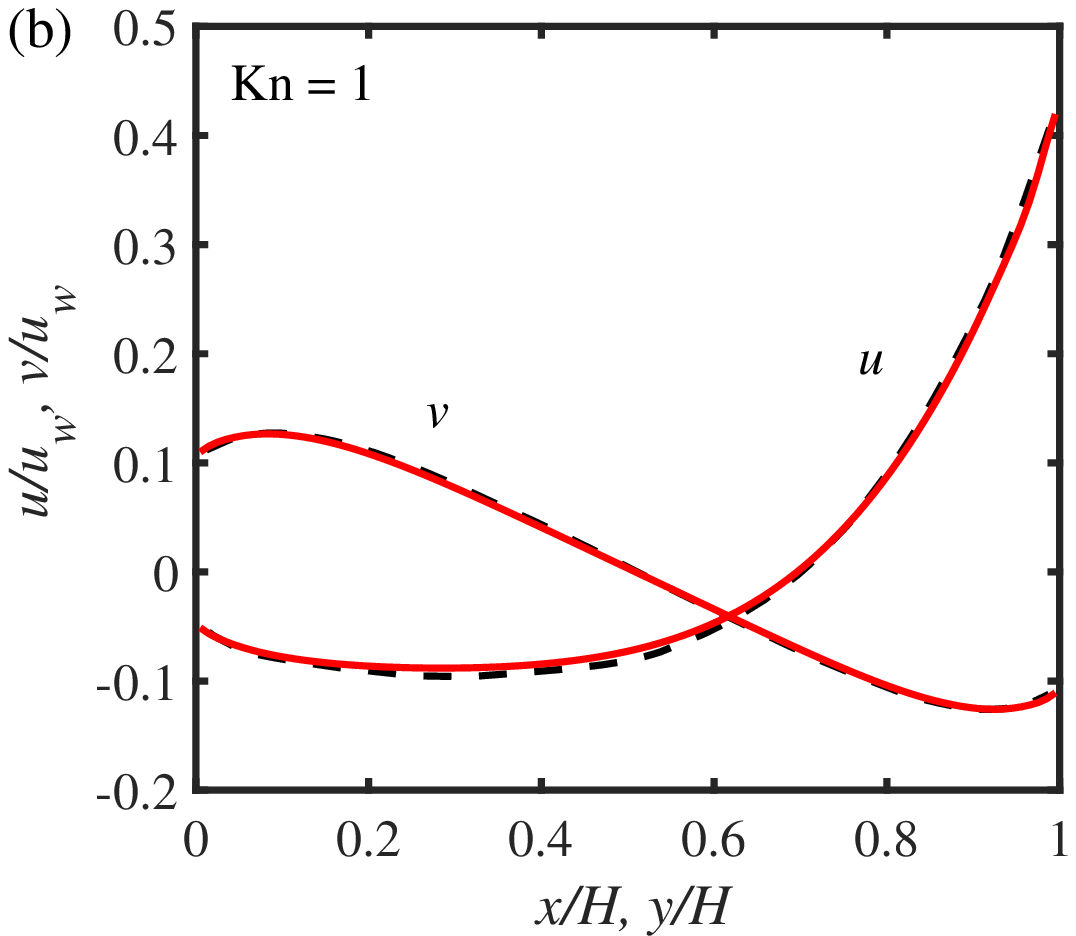}
        \label{fig:lid_driven_uv_Kn1}
    }%
    
    \subfigure{
        \includegraphics[height=4.5cm]{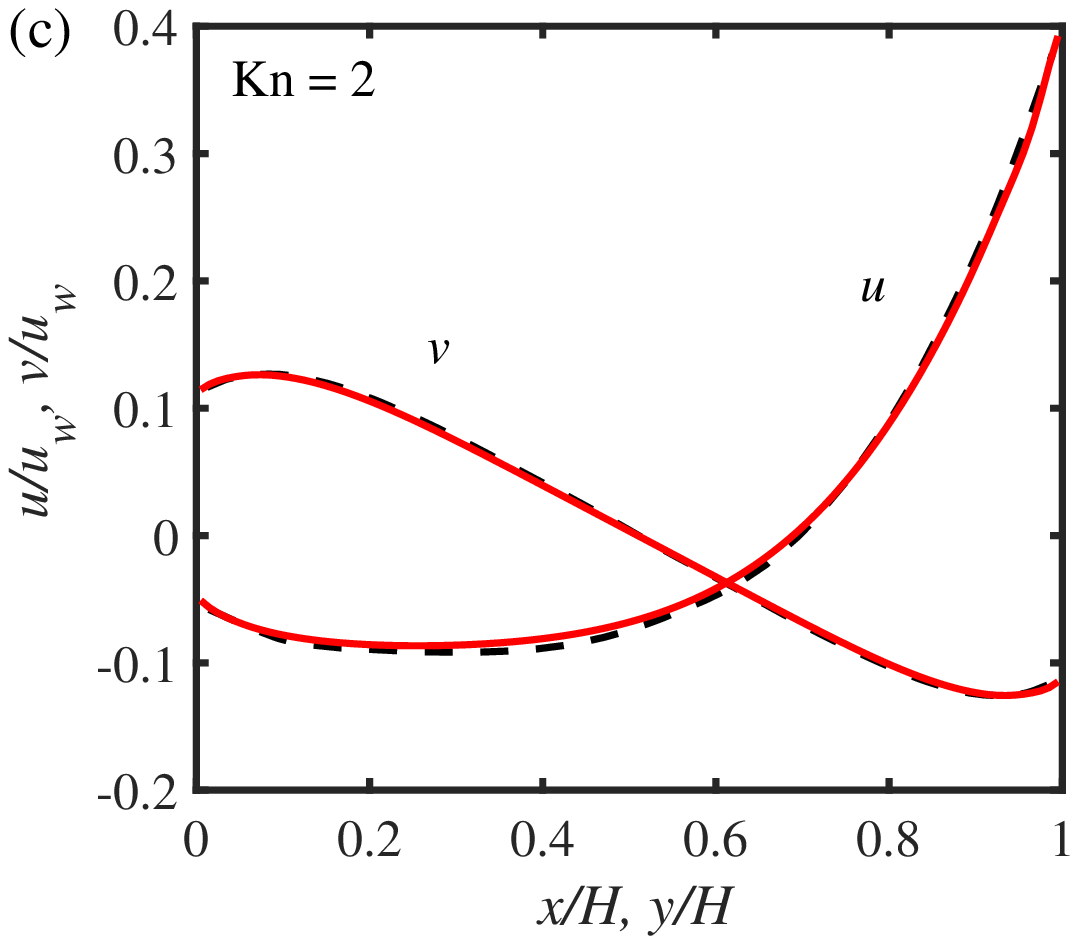}
        \label{fig:lid_driven_uv_Kn2}
    }%
    \subfigure{
        \includegraphics[height=4.5cm]{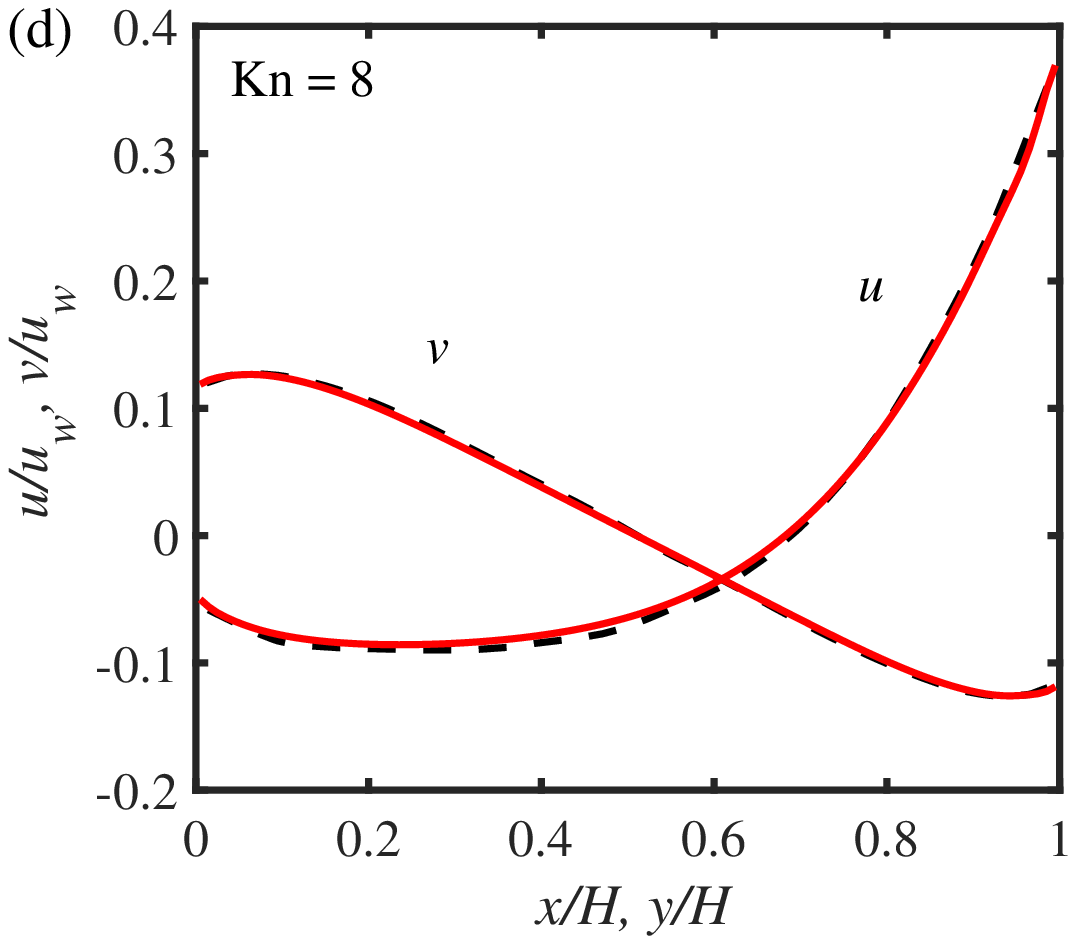}
        \label{fig:lid_driven_uv_Kn8}
    }%
    \caption{Microcavity flow: profiles of $u(y)|_{x=0.5H}$ and $v(x)|_{y=0.5H}$ for various Knudsen numbers. The red solid lines are our DVD-DVM simulations, and the black dashed lines are the reference data in \cite{John_2010}. For discrete directions and velocities, we set $N=30$ and $\{\bm l_m = (\cos\frac{(2m-1)\pi}{60}, \sin\frac{(2m-1)\pi}{60})\}_{m=1}^{30}$ while $\xi_k = 0.3k -8.85$ for $k=1,...,60$.
    }\label{fig:lid_driven_uv}
\end{figure}

Fig.~\ref{fig:lid_driven_streamlines} depicts the streamlines and the flow vector field for microcavity flows with various ${\rm Kn}$.
A bulk vortex is clearly observed and the streamlines are almost axisymmetric about the horizontal center $x=0.5H$.
As Kn increases, the height ($y$-value) of the vortex center reduces. These features were also presented in previous works \cite{Naris2005,Guo_2013}. Fig.~\ref{fig:lid_driven_uv} gives a comparison of the velocity profiles $\bm U = (u,v)^T$ across the cavity center with the reference data \cite{John_2010}. Both  $u(y)|_{x=0.5H}$ and $v(x)|_{y=0.5H}$ are plotted together for each $\rm Kn$.
It is seen that the DVD-DVM results are in good agreement with the reference data.

We next consider the flow with high Reynolds numbers $\rm Re$, where
\[{\rm Re} = \frac{u_w H}{\theta_0 \tau}.\]
In this case, we set ${\rm Re} = 1000$ by taking $u_w = 0.2$, $H=1$, $\theta_0 = 1$, and $\tau = 2\times 10^{-4}$.
This set of parameters characterises 
a nearly incompressible flow.
For the DVD-DVM, we set $N=6$ and the directions $\bm l_m = \left(\cos\frac{(2m-1)\pi}{12}, \sin\frac{(2m-1)\pi}{12} \right)^T$.
The SSP2 scheme is adopted with the bounce-back boundary condition for no-slip walls. The discrete velocity nodes in each direction are taken as $\xi_k=k-8.5$ for $k=1,...,16$.

\begin{figure}[htbp]
    \centering
    \subfigure{
        \includegraphics[height=5.5cm]{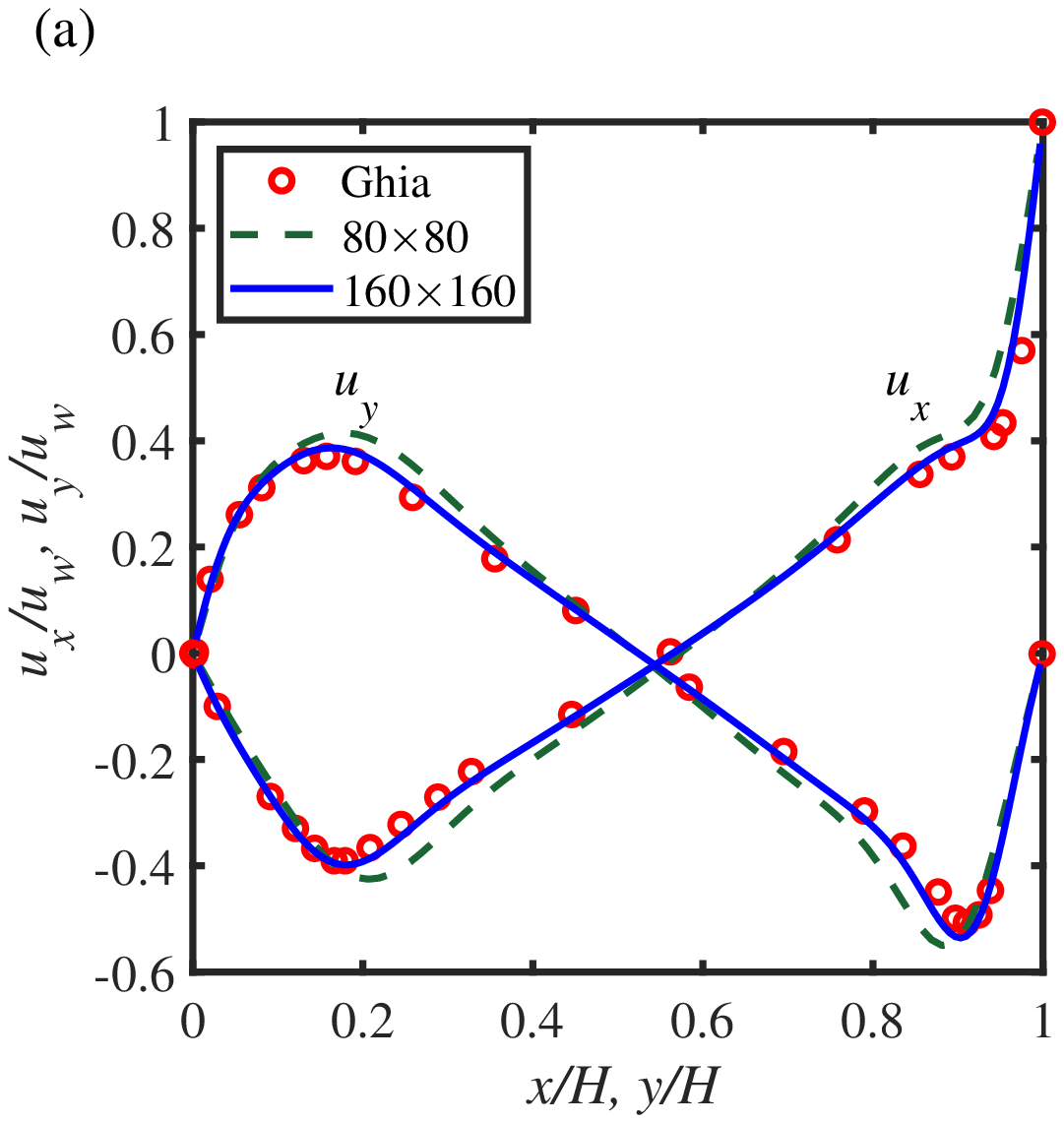}
    }%
    \subfigure{
        \includegraphics[height=5.5cm]{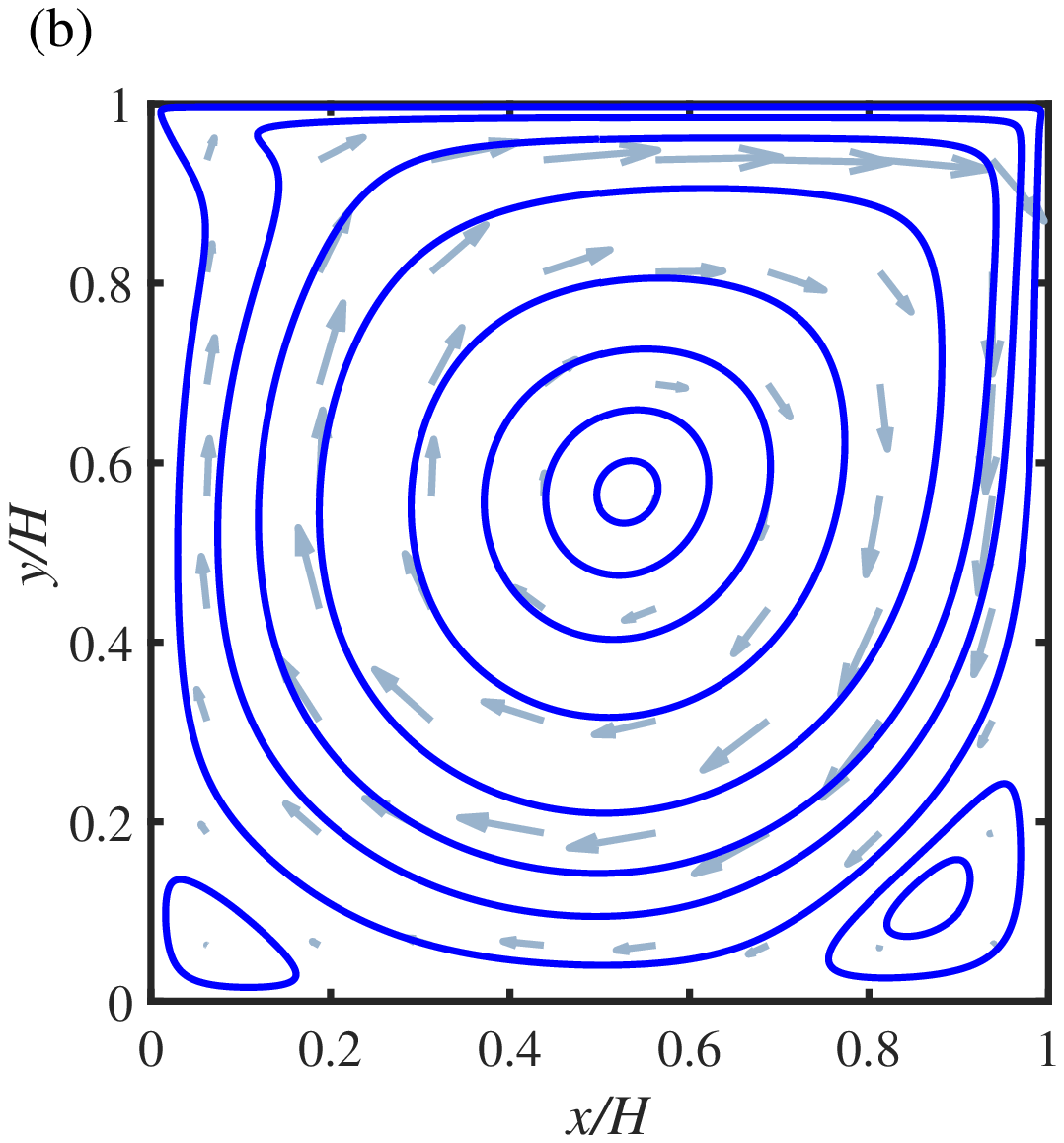}
    }%
    \caption{Lid-driven cavity flow: (a) profiles of $u(y)|_{x=0.5H}$ and $v(x)|_{y=0.5H}$ and (b) the streamlines for ${\rm Re} = 1000$. Red circles are benchmark data \cite{Ghia_1982}.
        The lines are the DVD-DVM results.
        The green dashed lines are from a $80\times80$ uniform grid and the blue solid lines are from a $160\times160$ uniform grid. We set $N=6$ and choose $\{\bm l_m\}_{m=1}^6$ as above. The discrete velocities in each direction are $\xi_k=k-8.5$ for $k=1,...,16$. The streamlines are based on data from the $160\times 160$ uniform grid.
    }\label{fig:lid_driven_Re1000_uv}
\end{figure}

Fig.~\ref{fig:lid_driven_Re1000_uv} shows the steady-state velocity profiles across the cavity center. The benchmark data are from \cite{Ghia_1982}. The physical domain $[0,H]^2$ is discretized to uniform cells. It is seen that when the uniform grids get finer (from $80\times80$ to $160\times160$), the simulation results become more accurate and well captures the highly nonlinear boundary profiles.

\section{Conclusions}
\label{sec:conclusions}

In this article, we have proposed a discrete-velocity-direction model (DVDM) based on the BGK equation with the internal molecular degrees of freedom. 
Assuming that the molecule velocity is restricted to a few prescribed directions but the velocity magnitude is still continuous, a semi-continuous DVDM is obtained, where the local discrete equilibrium in each direction is derived by the minimum entropy principle subject to the conservation laws. A key feature of the new model is the introduction of the weight function $|\xi|^{D-1}$ in the evaluation of the macroscopic fluid quantities. 

This DVDM can be combined with various treatments of 1-D velocity distribution functions to develop multidimensional spatial-time approximations of the original BGK equation. Specifically, three spatial-time DVD-submodels are derived by incorporating the discrete-velocity model (DVM), the 1-D Gaussian-EQMOM and a Hermite spectral method (HSM). We remark that the DVD-DVM allows radially-positioned discrete velocity nodes, whereas the DVD-EQMOM and DVD-HSM can be regarded as alternative multidimensional versions of EQMOM and HSM, respectively.

The feasibility of three spatial-time models have been verified numerically. 
For the numerical tests, the DVD-DVM and DVD-HSM are discretized with the second-order implicit-explicit Runge-Kutta scheme, while only the first-order upwind scheme is used for the DVD-EQMOM. Two widely-used limiting gas-solid boundary conditions, including the diffuse-scattering law and the bounce-back rule, are properly specified for the DVD-DVM and DVD-EQMOM. Only the Neumann condition is applied for the DVD-HSM.
The numerical results for 1-D and 2-D Riemann problems, especially in both the hydrodynamic and rarefied limits, illustrate the ability of the DVDM submodels to capture flow discontinuities. Furthermore, the simulations of the planar Couette flow and lid-driven cavity flow agree reasonably well with the benchmark data in a wide range of flow regimes. 

The numerical tests suggest that the DVD-DVM should be used for the rarefied flows. On the other hand, our numerical results are just preliminary. 
Better results are expected by using higher-order numerical schemes for spatial-time models or by enlarging the order $M$ of the DVD-EQMOM. These and the simulation of 3-D flows are our ongoing projects.

\section{Acknowledgments}
\label{sec:acknowledgments}

This work is supported by the National Key Research and Development Program of China (Grant no. 2021YFA0719200) and the National Natural Science Foundation of China (Grant no. 51906122 and 12071246).

\end{document}